\documentclass[preprint,showpacs,preprintnumbers,amsmath,amssymb, showkeys]{revtex4}

\usepackage{array}
\usepackage{booktabs}
\usepackage{tabu}
\usepackage{dcolumn}
\usepackage{amsmath}
\usepackage{amsfonts}
\usepackage{amssymb}
\usepackage{graphicx}
\usepackage{subfigure}
\usepackage{graphicx}
\usepackage{dcolumn}
\usepackage{bm}

\def\be{\begin{equation}}
\def\ee{\end{equation}}
\def\bea{\begin{eqnarray}}
\def\eea{\end{eqnarray}}
\def\f{\frac}
\def\n{\nonumber}
\def\l{\label}
\def\p{\phi}
\def\o{\over}
\def\R{\rho}
\def\pa{\partial}
\def\om{\omega}
\def\na{\nabla}
\def\P{\Phi}

\begin{document}

\title{Viscous warm inflation: Hamilton-Jacobi formalism}

\author{L. Akhtari}
\author{A. Mohammadi}
\email{abolhassanm@gmail.com}
\author{K. Sayar}
\author{Kh. Saaidi}

\affiliation{
Department of Physics, Faculty of Science, University of Kurdistan, Sanandaj, Iran.\\}
\date{\today}

\def\be{\begin{equation}}
\def\ee{\end{equation}}
\def\bea{\begin{eqnarray}}
\def\eea{\end{eqnarray}}
\def\f{\frac}
\def\n{\nonumber}
\def\l{\label}
\def\p{\phi}
\def\o{\over}
\def\R{\rho}
\def\pa{\partial}
\def\om{\omega}
\def\na{\nabla}
\def\P{\Phi}

\begin{abstract}
Using Hamilton-Jacobi formalism, the scenario of warm inflation with viscous pressure is considered. The formalism gives a way of computing the slow-rolling parameter without extra approximation, and it is well-known as a powerful method in cold inflation. The model is studied in detail for three different cases of the dissipation and bulk viscous pressure coefficients. In the first case where both coefficients are taken as constant, it is shown that the case could not portray warm inflationary scenario compatible with observational data even it is possible to restrict the model parameters. For other cases, the results shows that the model could properly predicts the perturbation parameters in which they stay in perfect agreement with Planck data. As a further argument, $r-n_s$ and $\alpha_s-n_s$ are drown that show the acquired result could stand in acceptable area expressing a compatibility with observational data.
\end{abstract}
\pacs{98.80.Cq}
\keywords{Warm inflation, Hamilton-Jacobian formalism, Viscous pressure}
\maketitle

\section{Introduction}
Since 1981, when Alan Guth proposed the idea of inflation for the first time \cite{Guth}, many types of inflationary models have been introduced \cite{Basset,Brandenberger}. The inflationary scenario predicts cosmological perturbations in which they play an essential role describing the universe structures, and at the same time the scenario could solve the drawback of big-bang theory such as horizon problem, flatness problem, and so on  \cite{Linde,Kolb,Weinberg,Mukhanov,Mukhanovbook}.  \\
A general classification of inflationary models is divided as cold and warm inflation. In cold inflation \cite{linde02,Baumann03,Liddle,abol,sheikh}, where the scalar field is dominant component and its interaction with other fields could be ignored, quantum perturbations are the source for cosmic microwaves background anisotropy and large scale structure as well. However, in warm inflation \cite{berera,berera2,taylor,oliveira,oliveira2,hall,gil}, scalar field, that is still the dominant component, interacts with other fields and decays through expansion into radiation and other fields \cite{oliveira2,hall}. Due to this phenomenon, the universe temperature does not reduced dramatically \cite{oliveira,oliveira2,hall,cid}, and there is no need for an extra phase of reheating to warm up the universe. Continuously producing radiation lets the universe smoothly enter the radiation-dominant era in order to have a successful big-bang nucleosynthesis \cite{oliveira2,hall,cid}.  One of the significant features of warm inflation is that temperature of the fluid during the inflation is larger than the Hubble parameter, $T>H$. Since thermal fluctuation and quantum fluctuation respectively depend on $T$ and $H$, during the warm inflation thermal fluctuation overcomes quantum fluctuation, and becomes the initial seeds for large scale structure of the universe.   \cite{herrera,herrera2,berera3,berera4}. Arisen density fluctuations due to thermal fluctuations have impact on scalar field through friction term in scalar field equation of motion \cite{berera3,berera5,berera6,hall2}. \\
Warm inflation with a self-interacting potential has been considered in \cite{gil02} for different types of the potential including monomial potential ($\propto \phi^n$). The dissipation coefficient is taken as a function of temperature and scalar field $C_\phi T^3/\phi^2$. The scenario is studied for different values of $n$. Plotting the $r-n_s$ diagram displays that, the quartic potential is compatible with observational data for $N=40$. However, the case with $n=2$ could not stand in acceptable range compared to the Planck data.  Dynamics of warm inflation including a self-interacting scalar field with potential $\lambda\phi^4$ has been studied in \cite{Panotopoulos} where the scenario investigated for strong and weak dissipative regime. The dissipation coefficient is taken as a function of time and they could successfully constrain the free parameters of the model using the Planck data in which the $r-n_s$ diagram stands in acceptable range. In \cite{Benetti}, the scenario of warm inflation has been investigated for different types of the potential including chaotic quartic potential, for two different types of dissipation coefficient as $T^3/\phi^2$ and $T$. Their result determined that in weak dissipative regime, the $r-n_s$ and $\alpha_s-n_s$ diagrams could be placed better for the second type of $\Gamma$. For strong dissipative regime the results are not as good as weak dissipative regime.\\
The produced particles from decaying inflaton decay are usually assumed as radiation, however taking into account the production of other particles with mass could change the dynamics by generalizing the fluid pressure in two ways \cite{delCampo}: I) the hydrodynamic, equilibrium pressure changes from $p=\rho/3$ to $p=(\gamma-1)\rho$, where $\rho$ is the energy-density of matter-radiation and $\gamma$ is adiabatic index standing in $1 \leq \gamma \leq 2$. II) rising non-equilibrium viscous pressure $\Pi$ in two different mechanisms: i) interparticle interaction \cite{Landau,Huang}; ii) particle decay within the fluid \cite{Zeldovich,Barrow,Zimdahl,Zimdahl2}. \\
Warm inflation with a viscous pressure has been studied in \cite{delCampo} where a chaotic potential $m^2\phi^2$ has been introduced for the scalar field. The situation is investigated for different kinds of dissipation coefficient and bulk viscosity. Using the observational data, the authors could constraint the free parameters of the model, however it seems that the running of scalar spectral index for the obtained free parameters could not stand in acceptable range. In \cite{sharif}, intermediate viscous warm inflation has been explore by assuming an anisotropic universe describing by BI model. The model was considered for different types of $\Gamma$ and $\xi$. Although the $r-n_s$ and $\alpha_s-n_s$ could be plotted perfectly, the obtained scalar spectral index stays in observational range only for small number of e-folds.  \\
Considering warm inflation by using Hamilton-Jacobi formalism, where instead of the potential the Hubble parameter is given as a function of scalar field \cite{salopek,liddle,kinney,guo,aghamohammadi,saaidi,sheikhahmadi}, is the main goal of the present work. Besides, the presence of viscous pressure is assumed which is indicated by the usual fluid dynamics as $\Pi=-3\zeta H$ \cite{Landau,Huang} in which $\zeta$ stands for phenomenological coefficient of bulk viscosity. Applying the second law of thermodynamics comes to a positive definite expression for $\zeta$ which in general view relies on fluid energy density. \\
The paper is organized as follows: the general dynamical equations of the model are derived in Sec.II. The paper is restricted to strong dissipative regime in Sec.III, and all perturbation parameters are extracted in the regime and more detail is acquired carefully for typical examples in separate subsections. Finally, the result is summarized in Sec.IV.

\section{Preliminary}\label{Sec2}
We assume that the matter has two components as self-interacting scalar field with energy density $\rho_\phi = \dot\phi^2 / 2 + V(\phi)$ and pressure $ p_\phi = \dot\phi^2 / 2 - V(\phi)$, and an imperfect fluid with energy density $\rho$ and total pressure $p+\Pi$. The dynamical equation of the model is given by
\begin{equation}\label{Friedmann}
H^2 = {1 \over 3} (\rho_\phi + \rho),
\end{equation}
where a spatially flat FLRW metric has been chosen. Since the scalar field interacts with the the other fields and it decays with rate $\Gamma$ into the fluid, the conservation equations are modified as follows
\begin{equation}
\dot\rho_\phi + 3H(\rho_\phi + p_\phi) = -\Gamma \dot\phi^2,
\end{equation}
\begin{equation}
\dot{\rho}+ 3H(\rho+p+\Pi) = \Gamma \dot{\phi}^2,
\end{equation}
so that $\Gamma$ is the dissipation coefficient that in general depends on $\phi$, and by the second law of the thermodynamics should be positive \cite{delCampo}. \\
The continuity equation of scalar field could be rewritten as follows
\begin{equation*}
\ddot\phi +(3H+\Gamma)\dot\phi + V'(\phi)=0,
\end{equation*}
which prime denotes derivative with respect to $\phi$. The equation is known as the scalar field equation of motion as well. A quasi-stable decay of scalar field follows by the condition $\dot\rho \ll 3H(\gamma\rho+\Pi), \Gamma \dot\phi^2$, then the radiation energy density could be estimated as
\begin{equation}\label{radiation}
\rho = \gamma^{-1} \big( Q\dot\phi^2 - \Pi \big).
\end{equation}
in which the parameter $Q$ is defined as $Q \equiv {\Gamma / 3H}$ that describe the quality of dissipation by distinguishing weak and strong dissipative regimes respectively corresponded to $Q \ll 1$ and $Q \gg 1$. Applying this condition, the second Freidmann equation is approximated as
\begin{equation}\label{Friedmann2}
\dot{H} = -{1 \over 2} \; \big( 1+Q  \big) \dot\phi^2.
\end{equation}
During the inflation the scalar field is still the dominant energy density. From the second Friedmann equation (\ref{Friedmann2}), the time derivative of scalar field is obtained as a function of scalar field
\begin{equation}\label{phidot}
\dot\phi = -{2 \over (1+Q)}\; H'(\phi).
\end{equation}
Substituting Eq.(\ref{phidot}) into Eq.(\ref{Friedmann}), and using the definition of $\rho_\phi$, the potential is obtained as
\begin{equation}\label{pot}
V(\phi) = 3H^2(\phi) - {2 \over (1+Q)^2}H'^2(\phi),
\end{equation}
where the last two equations are known as Hamilton-Jacobi equations. \\
The slow-rolling approximation could be perfectly described by the slow-rolling parameters. The first and maybe the most important slow-rolling parameter is $\epsilon$ which is given by
\begin{equation}\label{epsilon}
\epsilon \equiv -{\dot{H} \over H^2} = {2 \over (1+Q)}\; {H'^2(\phi) \over H^2(\phi)}.
\end{equation}
Above expression could be used to rewrite the fluid energy density as
\begin{equation}\label{fenergy}
\rho = {1 \over \gamma}\left( {2Q \over 3(1+Q)}\; \epsilon \rho_\phi - \Pi \right).
\end{equation}
Therefore, at the end of inflation, when $\epsilon=1$, one could find that $\rho= \left( 2 \rho_\phi/3 - \Pi \right)/\gamma$ for for the case $Q \gg 1$. \\
The amount of inflation is expressed by number of e-fold read by
\begin{equation}\label{efold}
 N = \int_{\phi_e}^{\phi}{1 \over 2} (1+Q){H(\phi) \over H'(\phi)} d\phi .
\end{equation}
The second important slow-roll parameter is given by $\epsilon_2  \equiv -{\ddot{H} / H\dot{H}}$ \cite{Kolb,Lyth,Linde2}, which leads to the following parameters
\begin{equation}\label{eta}
\eta ={4 \over (1+Q)}\; {H''(\phi) \over H(\phi)}, \qquad, \quad \beta={2  \over (1+Q)}\; {\Gamma' H' \over \Gamma H}.
\end{equation}
Moreover, there are some other parameters which would be useful in future calculation, and we prefer to introduce them here
\begin{equation}\label{srp}
\sigma = {2 \over (1+Q)}\; {H''' \over  H'}, \qquad, \quad \delta={4  \over (1+Q)^2}\; {\Gamma'' H'^2 \over \Gamma H^2}.
\end{equation}
The validity of any inflationary models should be checked in comparison to the observational data. Following \cite{delCampo,sharif,jawad}, the amplitude of scalar perturbation is read as
\begin{equation}\label{amplitud}
 \mathcal{P}_s =  \big(64  \pi^{2} \big) \; {\exp\big[ -2\chi(\phi) \big] \over \big[V'(\phi)\big]^2} \; \delta\phi^2 ,
\end{equation}
in which the parameter $\chi(\phi)$ is defined by \cite{delCampo,sharif,jawad}
\begin{eqnarray*}
\chi(\phi) & = &   - \int \left( \dfrac{\Gamma'}{3H + \Gamma} + \dfrac{3}{8 G(\phi)}\dfrac{\Gamma + 6H}{(\Gamma + 3H)^2} \hspace{0.5cm}  \right.  \\
   &     &   \quad \left. \times \left[ \Gamma + 4H - \Big[(\gamma - 1) + \Pi \dfrac{\zeta_{,\rho}}{\zeta} \Big] \dfrac{\Gamma' V'}{3 \gamma H (3H + \Gamma) } \right] \dfrac{V'}{V} \right) d\phi,
\end{eqnarray*}
and
\begin{equation*}
G(\phi) = 1 - {1 \over 8H^2} \left( 2\gamma\rho + 3\Pi + {\gamma\rho + \Pi \over \gamma} \Big[ {\xi_{,\rho} \over \xi}\Pi - 1 \Big] \right).
\end{equation*}
The scalar spectral index is derived by taking derivative of $\mathcal{P}_s$, so that
\begin{equation}\label{ns}
n_s - 1 = {d\ln\big(\mathcal{P}_s \big) \over d\ln(k)},
\end{equation}
and by taking another derivative, one could find the running of scalar spectral index as
\begin{equation}\label{running}
\alpha_s ={d\big({n}_s \big) \over d\ln(k)}.
\end{equation}
Besides scalar perturbations, tensor perturbations is another prediction of inflationary scenario which is known as gravitational waves. Tensor perturbation is detected indirectly by measuring $r$ that is the tensor-to-scalar ratio, $r=\mathcal{P}_t / \mathcal{P}_s$ \cite{delCampo,Bhattacharya}. \\
Out of strong dissipative regime, the universe expansion speedily dilutes the radiation and particles which are produced by decaying the inflaton and heavy fields. Then, they have a low chance of having interaction and providing a considerable bulk viscosity. As a further argument it should be mentioned that, if $r$ is not big, then the hydrodynamic expression $\Pi = -3\xi H$ will no longer be valid (for more argument refer to \cite{delCampo,sharif,delcampo2}). Consequently, the work will be restricted to strong dissipative regime because it seems that it is the suitable regime to have viscosity. \\
In the following section, we are going to examine free parameters of the model by using Planck data. In this regards, a power-law function of scalar field is proposed for the Hubble parameter, $H(\phi)=H_0\phi^n$, where $n=0.5, 1, 2$ are our main interest (Note that generally the parameter $H_0$ should have the dimension of $M^{1-n}$). \\

\section{Strong Dissipative Regime}
In strong dissipative regime, $Q \gg 1$, the coefficient $(1+Q)$ that appears in the main dynamical equations could be approximated as $(1+Q) \simeq Q$. On the other side, in the warm inflationary scenario, instead of quantum perturbations, thermal perturbations produce density fluctuations. Therefore, this subject affects on scalar field in which there is $\delta\phi^2=k_FT_r/2\pi^2$ where $k_F=\sqrt{\Gamma H}$ and $T_r$ is the temperature of fluid \cite{berera2,taylor,herrera,herrera2,delCampo}. Substituting $\delta\phi^2$ in Eq.(\ref{amplitud}) comes to the amplitude of scalar perturbations for strong dissipative regime as
\begin{equation}\label{strongSP}
\mathcal{P}_s= \dfrac{8}{9} \; {  \Gamma^{1/2} \; T_r \over H^{3/2} \; H'^2} \; \exp[-2\bar\chi(\phi)],
\end{equation}
where $\bar\chi$ is computed $\chi$ for strong dissipative regime. Then, one could easily derive the scalar spectral index and its running as important perturbation parameters

\begin{equation}\label{strongns}
n_s = 1 + {3 \over 2}\; \epsilon + \eta - {1 \over2} \beta +{4 \over Q}{H' \over H} \; \bar\chi'(\phi),
\end{equation}
\begin{eqnarray}\label{strongrunning}
\alpha_s & = &   - \beta^{2} +{3 \over2}\beta\epsilon + {1 \over2}\delta + {5 \over4} \beta\eta + {3 \over 2}\epsilon^{2} - {3 \over2}\epsilon\eta - 2\epsilon\sigma \nonumber \\
  &  & + (2\beta -\eta)\epsilon \; {H \bar\chi'(\phi) \over H'} - {4 \over Q}\epsilon \bar\chi''.
\end{eqnarray}
The tensor-to-scalar parameter that indirectly measure the tensor perturbations in the regime is evaluated as
\begin{equation}\label{strongr}
r ={9 \over 16\pi^2 } { H^{7/2} H'^2 \over \Gamma^{1/2} T_r } \; \exp[2\bar\chi(\phi)] \coth\Big({k \over 2T}\Big).
\end{equation}
The extra temperature dependent term $\coth\big(k/2T\big)$ which appears through amplitude of tensor perturbations, is because production of tensor perturbations during inflation leads to stimulated emission in the thermal background of gravitational waves \cite{delCampo,Bhattacharya}(On the other side, according to \cite{taylor} tensor perturbation do not coupled strongly to the thermal background and therefore the gravitational waves are only produced by quantum fluctuations). Further detail requires an specific form of both dissipation coefficient $\Gamma$ and $\Pi$. The dissipation coefficient could be taken as a constant however in a general view it could depend on scalar field $\phi$. The bulk viscous pressure commonly is taken as $\Pi=-3\xi H$, in which $\xi$ denotes coefficient of bulk viscosity that could be constant or generally a function of fluid energy density. The situation will be studied in great detail for different examples of $\Gamma$ and $\xi$. \\

\subsection{Typical Example 1: $\Gamma := \Gamma_0$ and $\xi=\xi_0$}
\begin{figure}
 \centering
 \includegraphics[width=6cm]{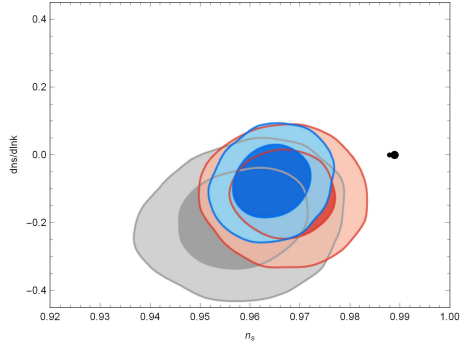}
 \caption{$\alpha_s-n_s$ diagram for $N=55-60$ and $n=0.5$.}\label{asns11}
\end{figure}

As the first example, we are about to consider the simplest case and take both the dissipation coefficient and bulk viscous coefficient as constant. Inflation ends when the slow-rolling parameter $\epsilon$ reaches one ($\ddot{a}=0$), and the final scalar field could be read from Eq.(\ref{epsilon}). Then, from Eq.(\ref{efold}), it could be concluded that the perturbations cross the horizon at the earlier times as the scalar field is
\begin{equation}\label{phii1}
  \phi_\ast = \phi_e \Big[ {N(2-n) \over n} +1 \Big]^{1 \over 2-n}, \qquad \phi_e = \left( \dfrac{6 n^2 H_{0}}{\Gamma_{0}} \right)^{1 \over 2-n}.
\end{equation}
Therefore, the non-vanishing slow-rolling parameter at this point could be estimated as
\begin{equation*}
\epsilon^\ast = {n \over n + (2-n)N}, \quad \eta^\ast = {2(n-1) \over n}\; \epsilon^\ast, \quad \sigma^\ast = {(n-2) \over 2n}\; \eta^\ast,
\end{equation*}
and since for the case $\Gamma$ is a constant, other slow-rolling parameters vanish $\beta^\ast=\delta^\ast=0$. Using some approximations that are reasonable in slow-rolling warm inflation, namely $H^2 \simeq \rho_\phi \gg \rho , \Pi$ (where it is assumed that the fluid energy density and the bulk viscosity are as same order; refer to \cite{cid}), the last term on the right hand side (r.h.s) of Eq.(\ref{ns}) could be approximated as
\begin{equation*}
{4 H' \over Q H} \; \bar{\chi} \simeq  - {3 \over 2} \epsilon^\ast
\end{equation*}
Substituting above relation in Eq.(\ref{strongns}), it is found out that for our interested value of $n$ the case could not predict the proper value for the scalar spectral index. Then, this case should be abandoned. This inconsistency with observation could be shown in $\alpha_s-n_s$ diagram as Fig.\ref{asns11} which is the best situation for the case. It is clear that the line stands far out of the acceptable range. For other choices of $n$ the situation gets even worse.\\
Note that, a different perspective could be applied that is explained in the following lines. In this point of view, one could try to restrict the parameters of the model by using the condition of warm inflation as $Q \gg 1$, $T>H$, $r^\ast < 0.11$ and $\mathcal{P}_s$ (the work we will do in the next sections). Doing so, one could properly constrain the parameter and also arrives at acceptable values for the scalar spectral index too. However, the major problem occurs as one depicts the energy densities of the scalar field and the fluid where it is discovered that the fluid energy density is bigger than the scalar field energy density, a clear contradiction with our fundamental assumption of inflation. Note that standing $n_s$ in the acceptable range is because of the fact that the fluid energy density becomes bigger than the scalar field energy density that affects $n_s$ through the term $4 H' \bar{\chi}'/QH$. Therefore, both perspectives result that this case is not acceptable and could not suitably describe the warm inflationary scenario.

\subsection{Typical Example 2: $\Gamma := \Gamma_0 \phi^m$ and $\xi=\xi_0$}
\begin{figure}
 \centering
 \includegraphics[width=6cm]{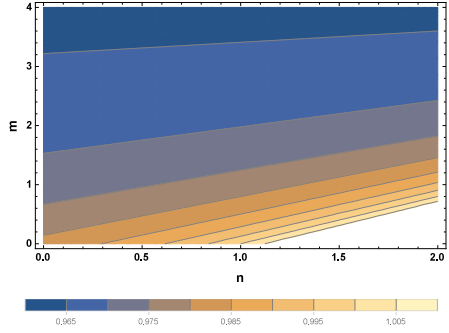}
 \caption{Contour plot of $n_s$ versus $n$ and $m$.}\label{ns2}
\end{figure}

\begin{figure}[h]
 \centering
 \subfigure[$n=0.5$]{\includegraphics[width=4.3cm]{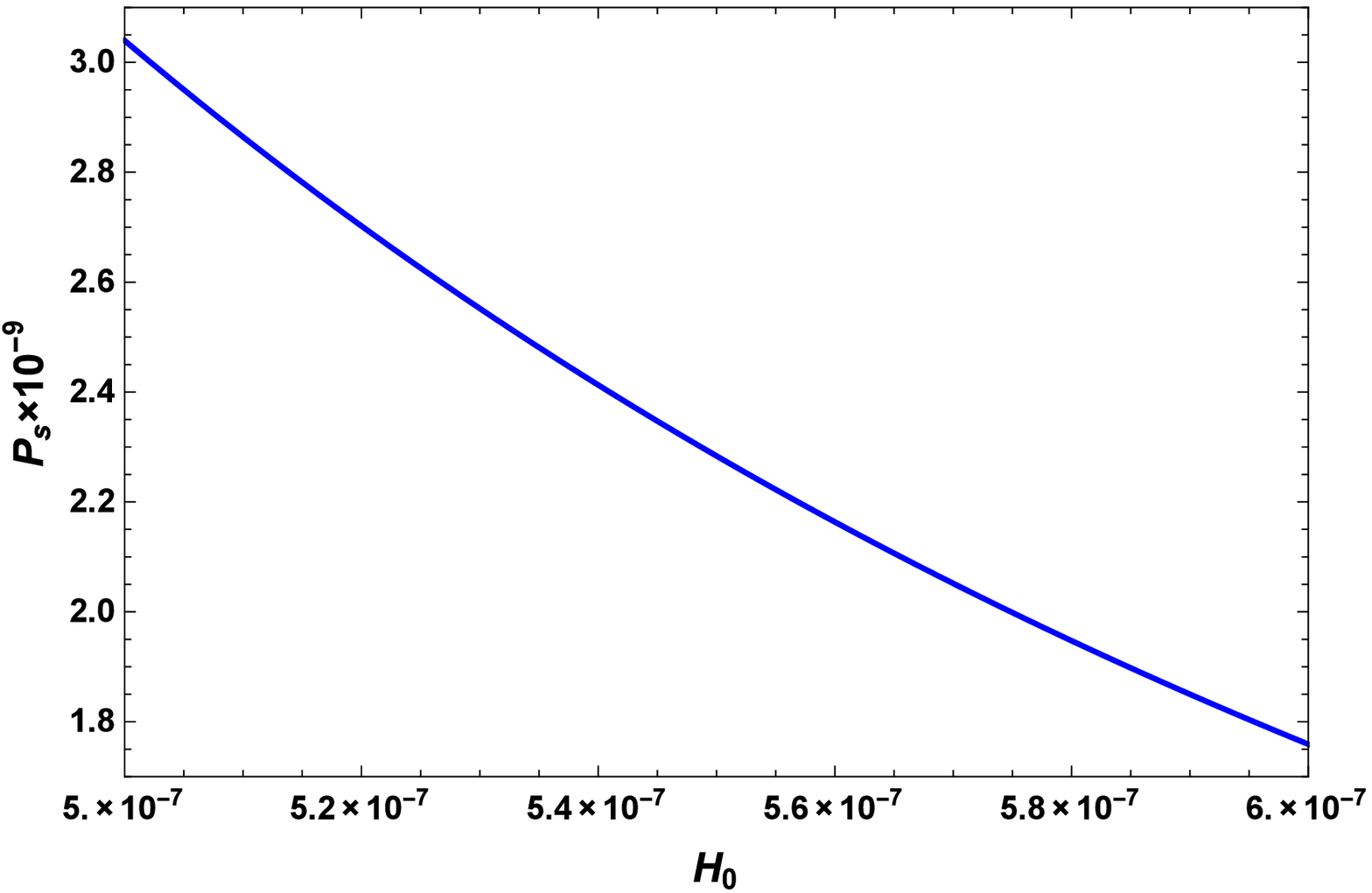}}
 \subfigure[$n=1$]{\includegraphics[width=4.3cm]{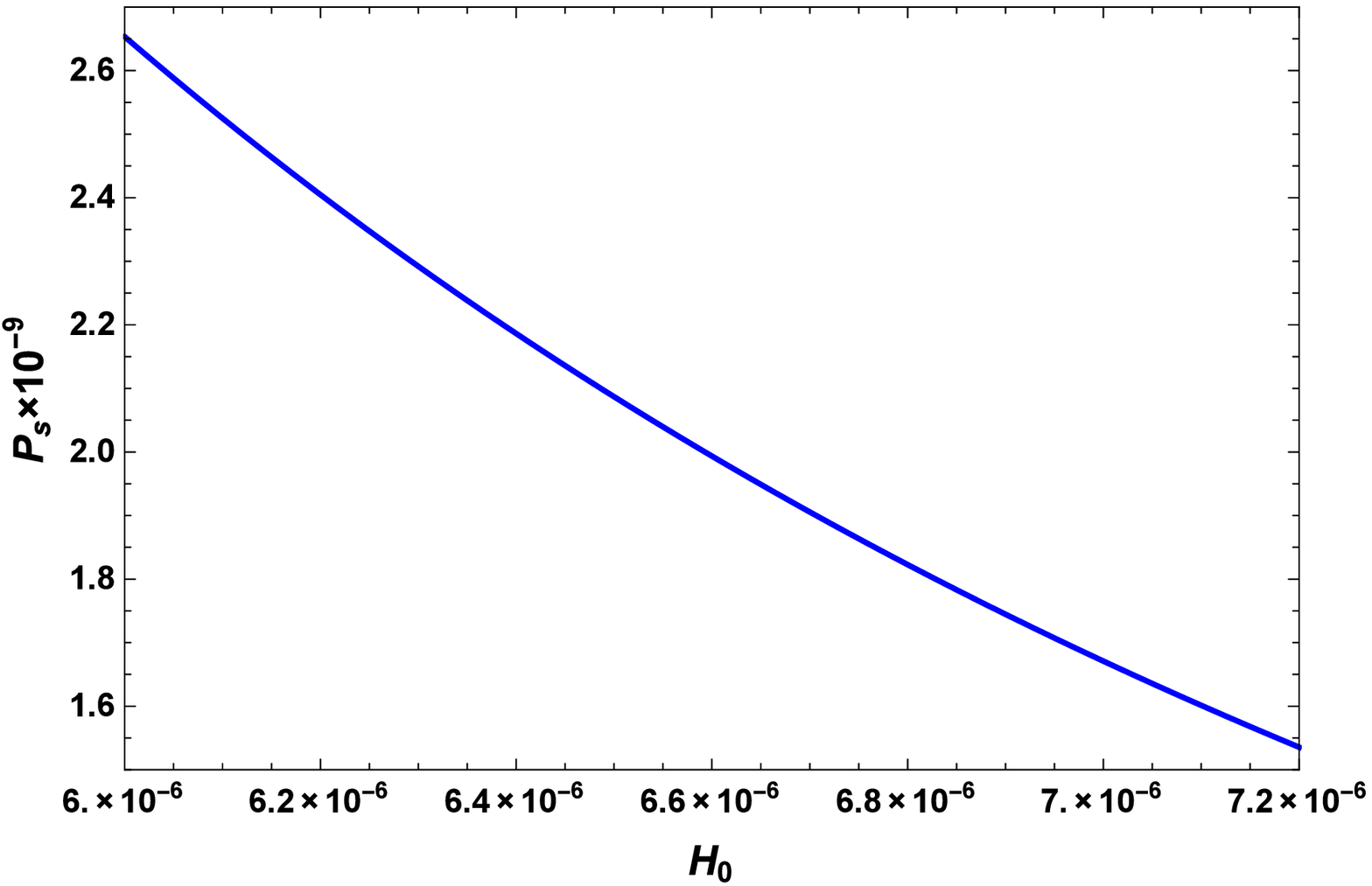}}
 \subfigure[$n=2$]{\includegraphics[width=4.3cm]{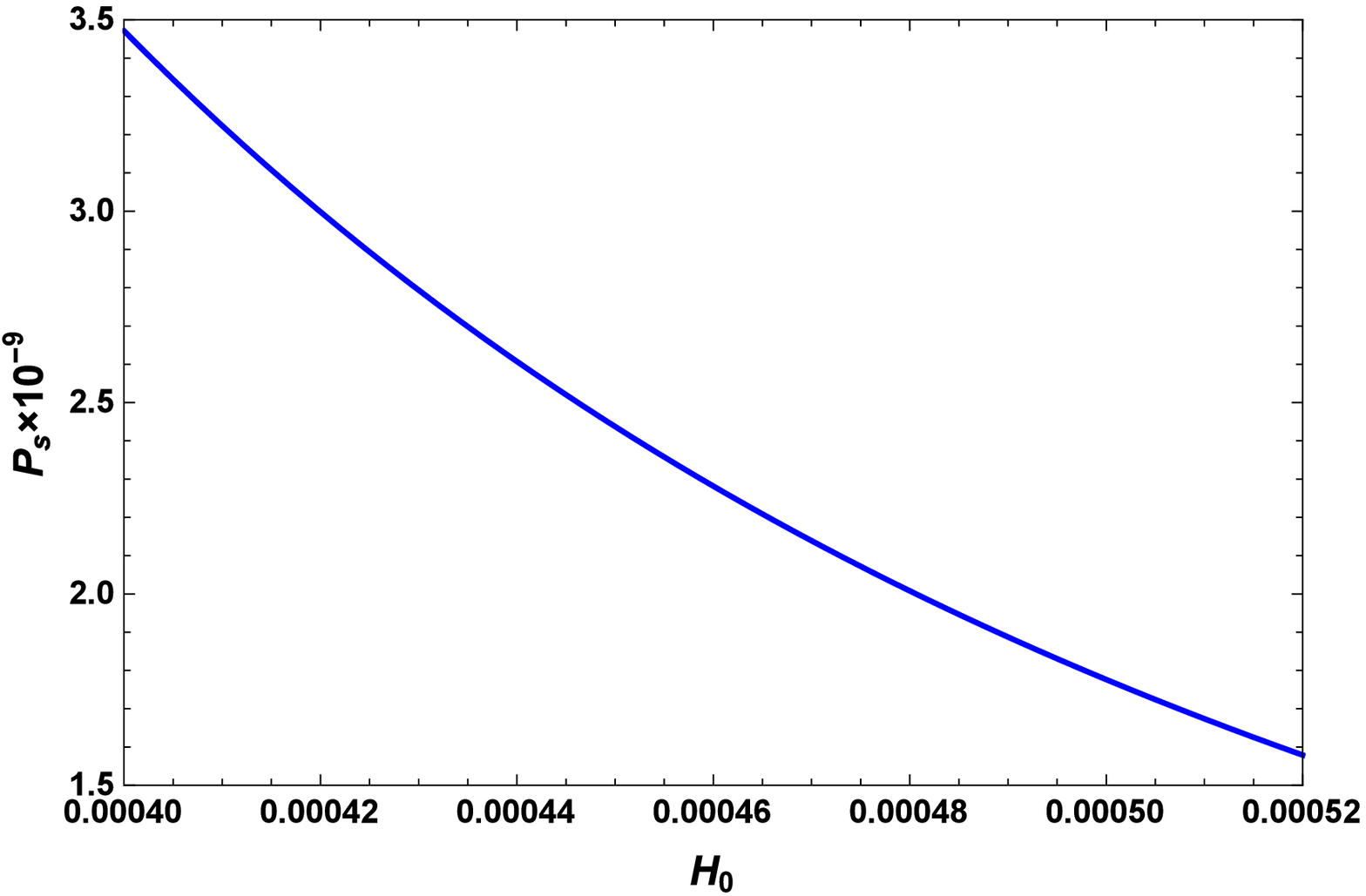}}
 \caption{The amplitude of scalar perturbation in term of the constant parameter $H_0$ is plotted for different values of $n$, where the constant parameters are: $\gamma=1.5$, $m=3$, $N=55$, $\theta=2\time 10^{10}$ $\xi_0=7 \times 10^{-14}$, $k=0.002$ and $T_r=T=2 \times 10^{-5}$. }\label{psG}
\end{figure}

\begin{figure}[h]
 \centering
 \subfigure[$n=0.5$]{\includegraphics[width=4.3cm]{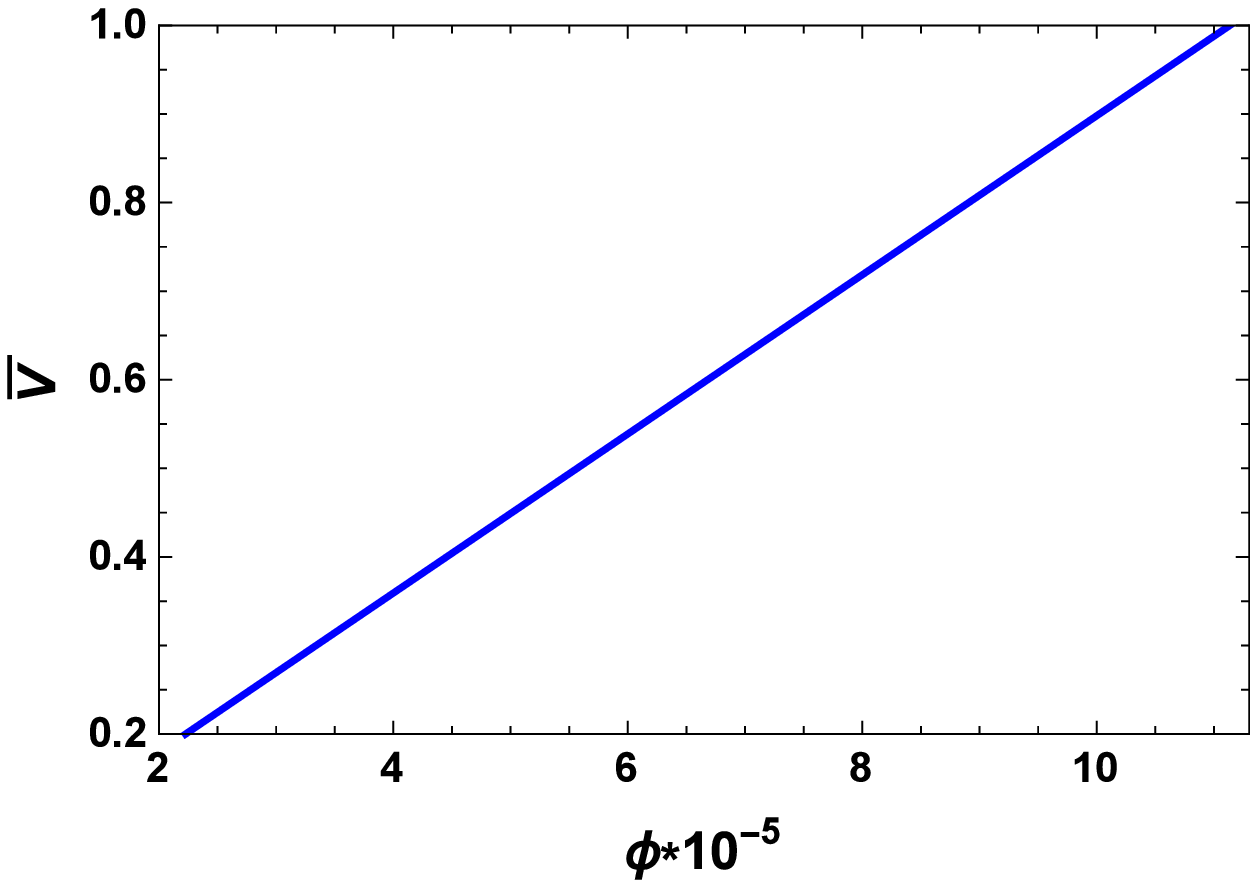}}
 \subfigure[$n=1$]{\includegraphics[width=4.3cm]{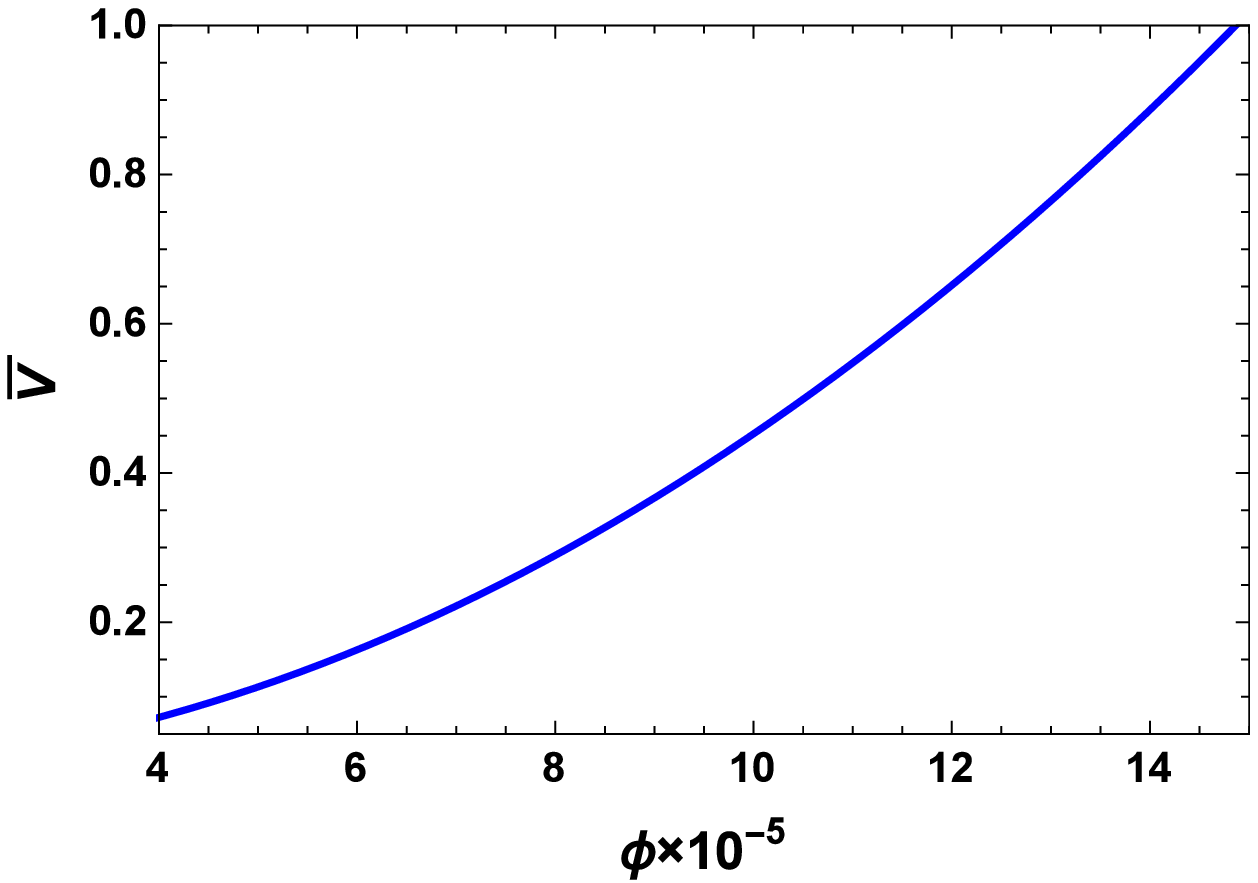}}
 \subfigure[$n=2$]{\includegraphics[width=4.3cm]{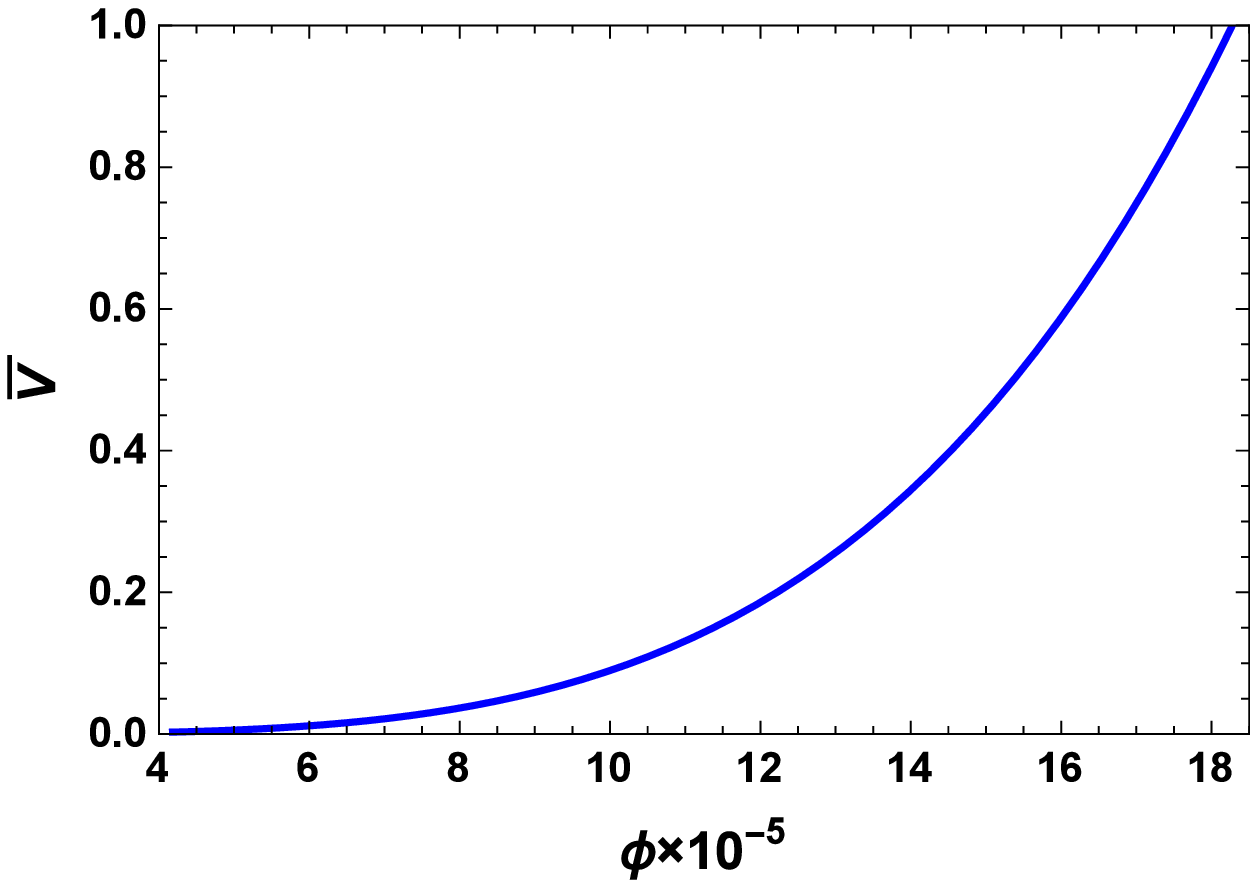}}
 \caption{The potential versus scalar field during the inflationary time where the constant parameters are taken as: $\gamma=1.5$, $m=3$, $N=55$, $\theta=2\time 10^{10}$ $\xi_0=7 \times 10^{-14}$, $k=0.002$ and $T_r=T=2 \times 10^{-5}$. The parameter $\bar{V}$ is defined by $\bar{V}=V/V_i$ where $V_i$ is the initial value of the potential.}\label{pot2}
\end{figure}

\begin{figure}[h]
 \centering
 \subfigure[$n=0.5$]{\includegraphics[width=4.3cm]{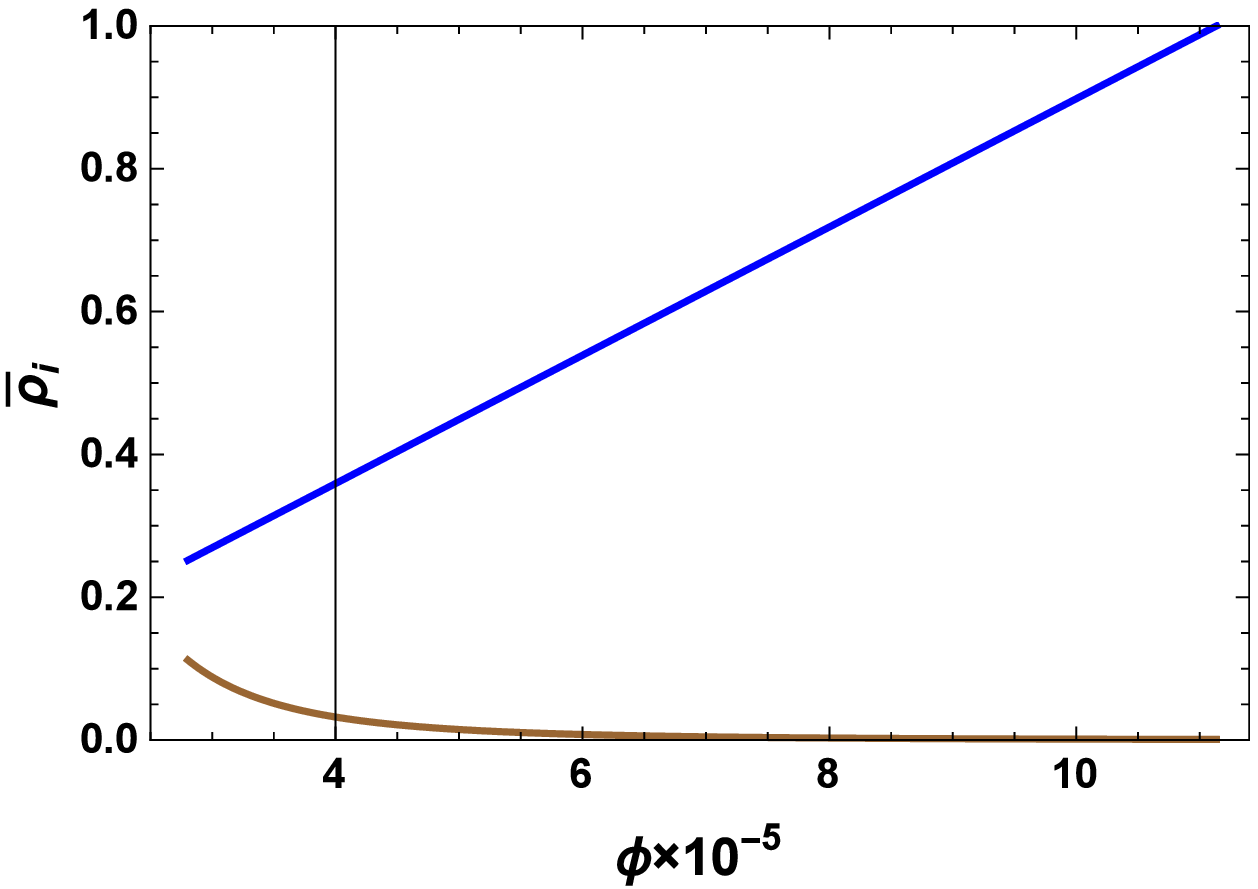}}
 \subfigure[$n=1$]{\includegraphics[width=4.3cm]{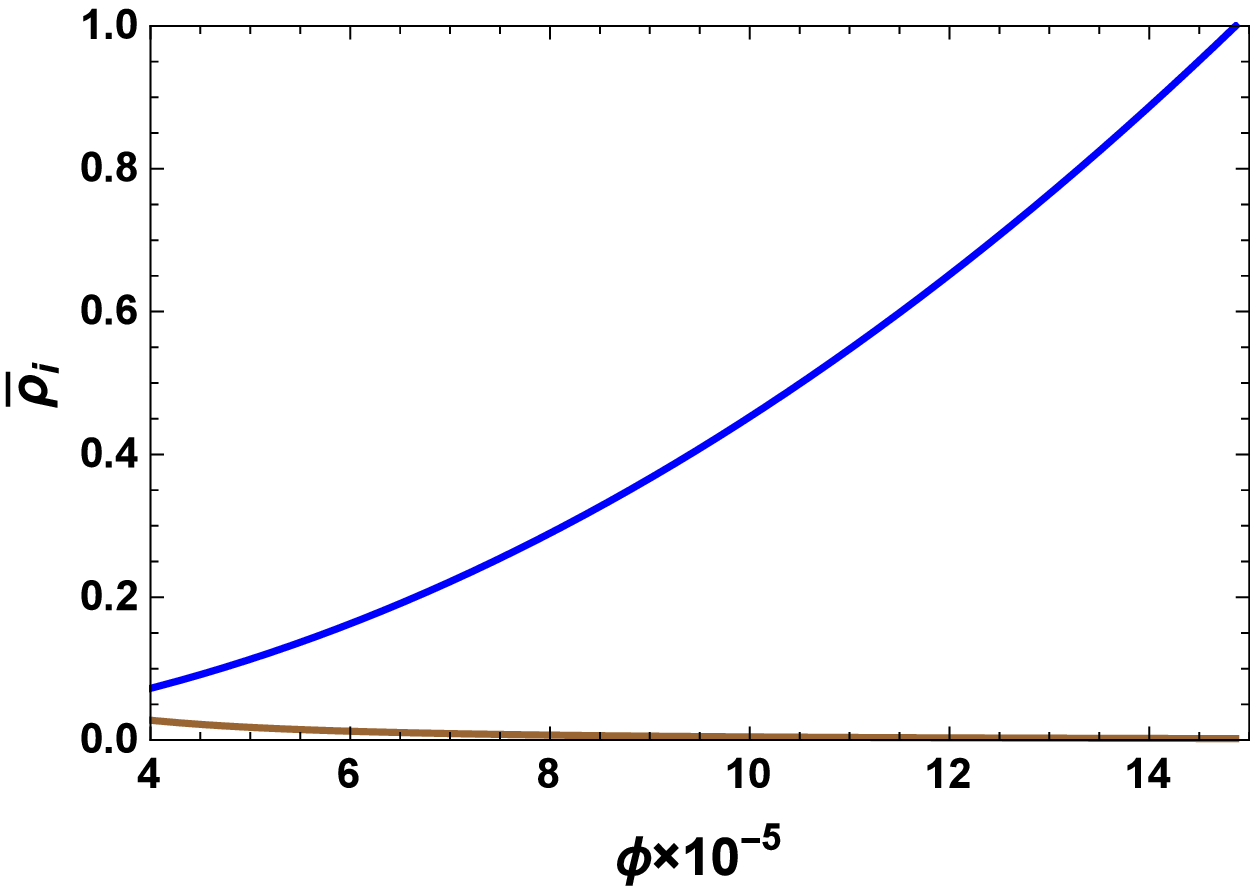}}
 \subfigure[$n=2$]{\includegraphics[width=4.3cm]{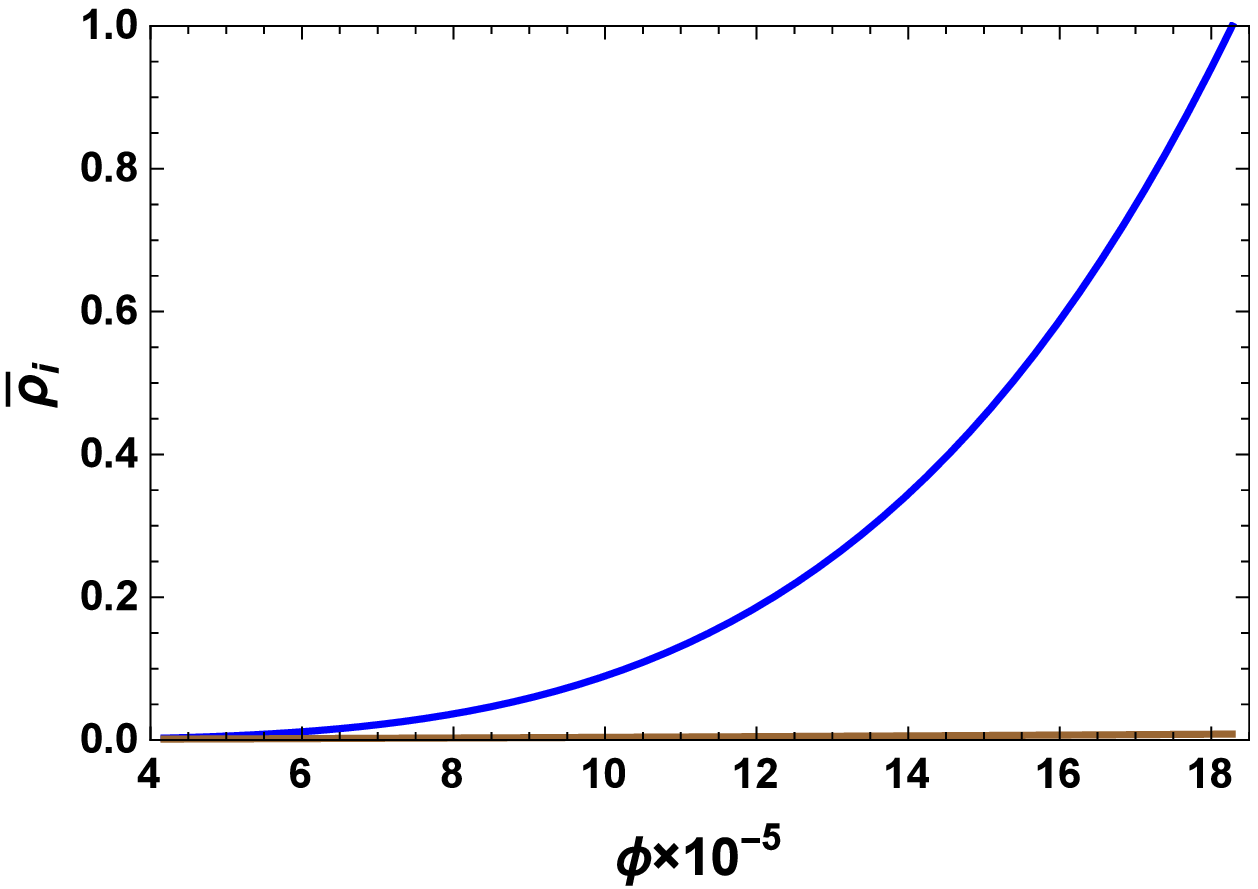}}
 \caption{The energy densities of scalar field (Blue line) and fluid (red line) versus scalar field during the inflationary times for the constant parameters: $\gamma=1.5$, $m=3$, $N=55$, $\theta=2\time 10^{10}$ $\xi_0=7 \times 10^{-14}$, $k=0.002$ and $T_r=T=2 \times 10^{-5}$. The parameter $\bar{\rho}_i$ is defined by $\bar{\rho}_i=\rho_i/\rho_0$ (the subscript "i" stands for scalar field and radiation) where $\rho$ is the initial value of the scalar field energy density.}\label{energya}
\end{figure}

\begin{figure}[h]
 \centering
 \subfigure[$r-n_s (n=0.5)$]{\includegraphics[width=4.3cm]{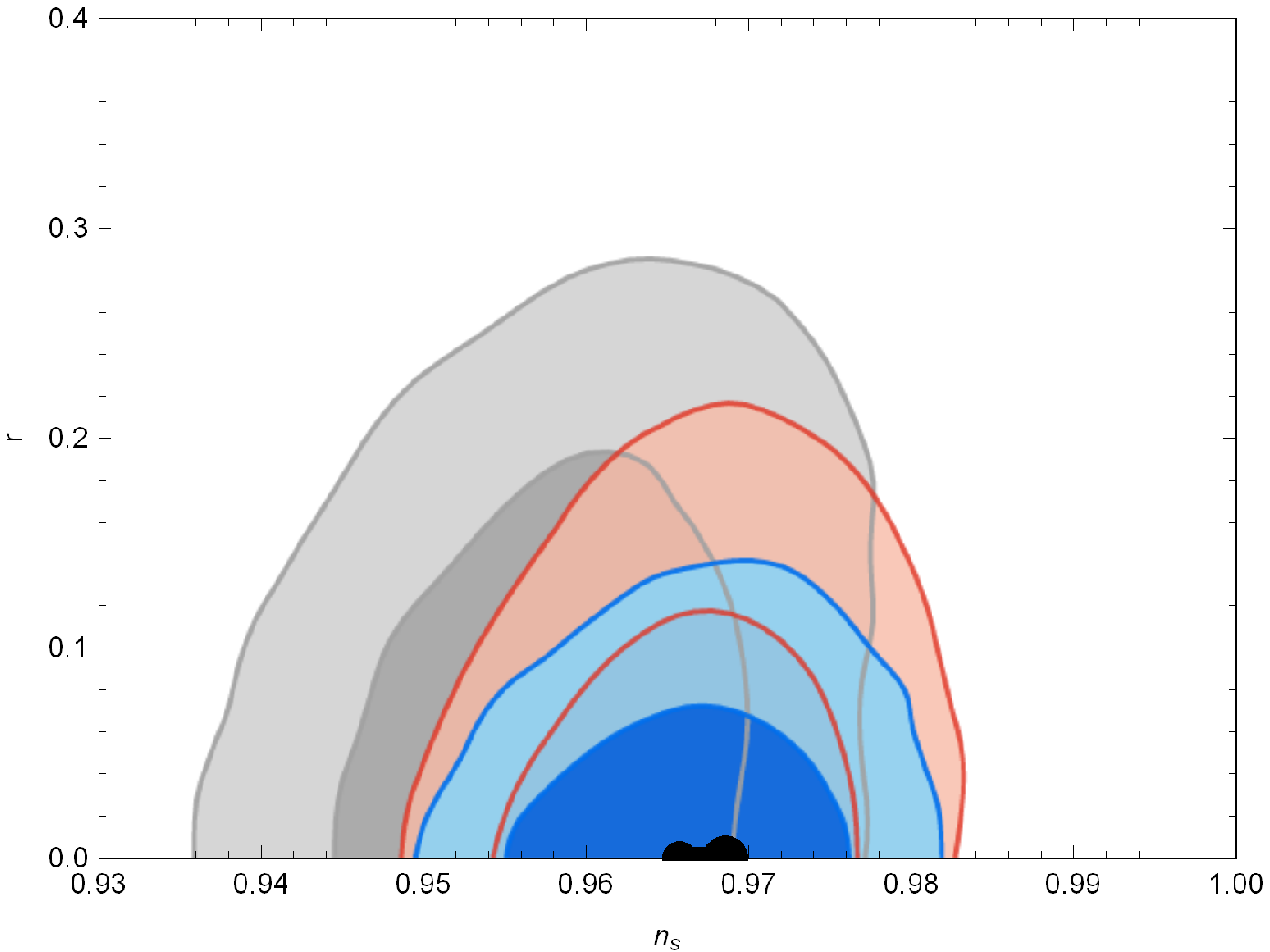}}
 \subfigure[$r-n_s (n=1)$]{\includegraphics[width=4.3cm]{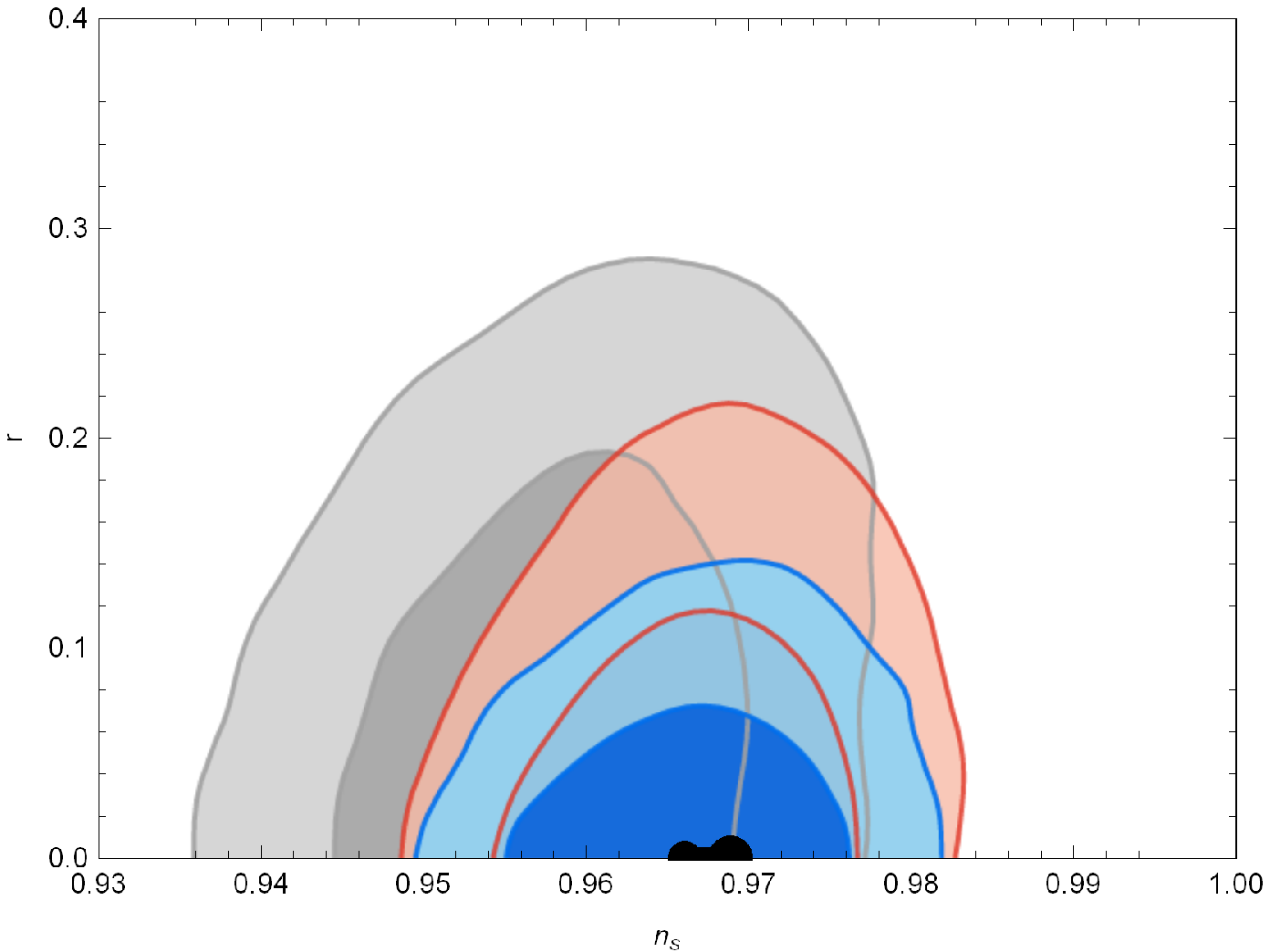}}
 \subfigure[$r-n_s (n=2)$]{\includegraphics[width=4.3cm]{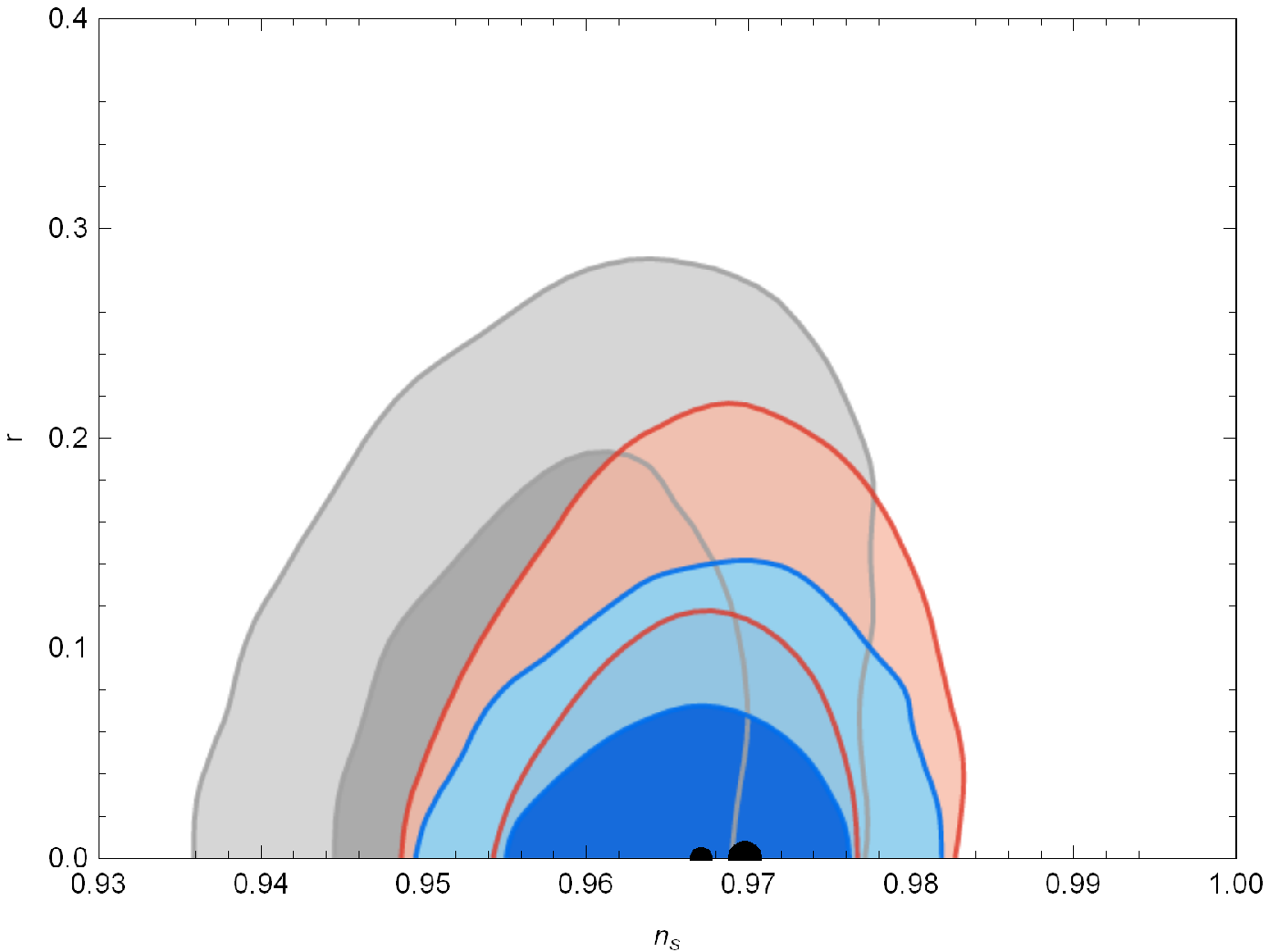}}
 \subfigure[$\alpha_s-n_s (n=0.5)$]{\includegraphics[width=4.3cm]{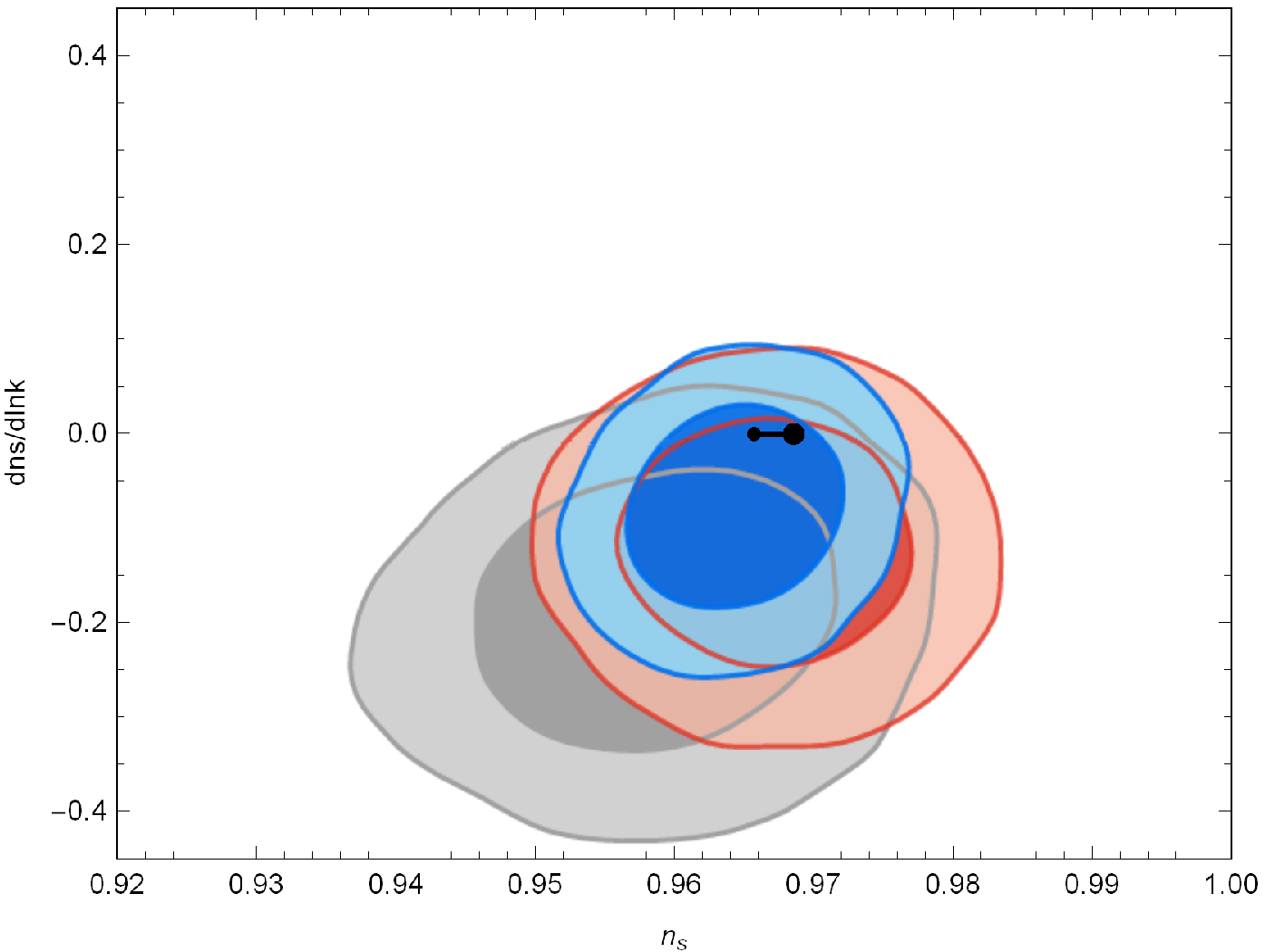}}
 \subfigure[$\alpha_s-n_s (n=1)$]{\includegraphics[width=4.3cm]{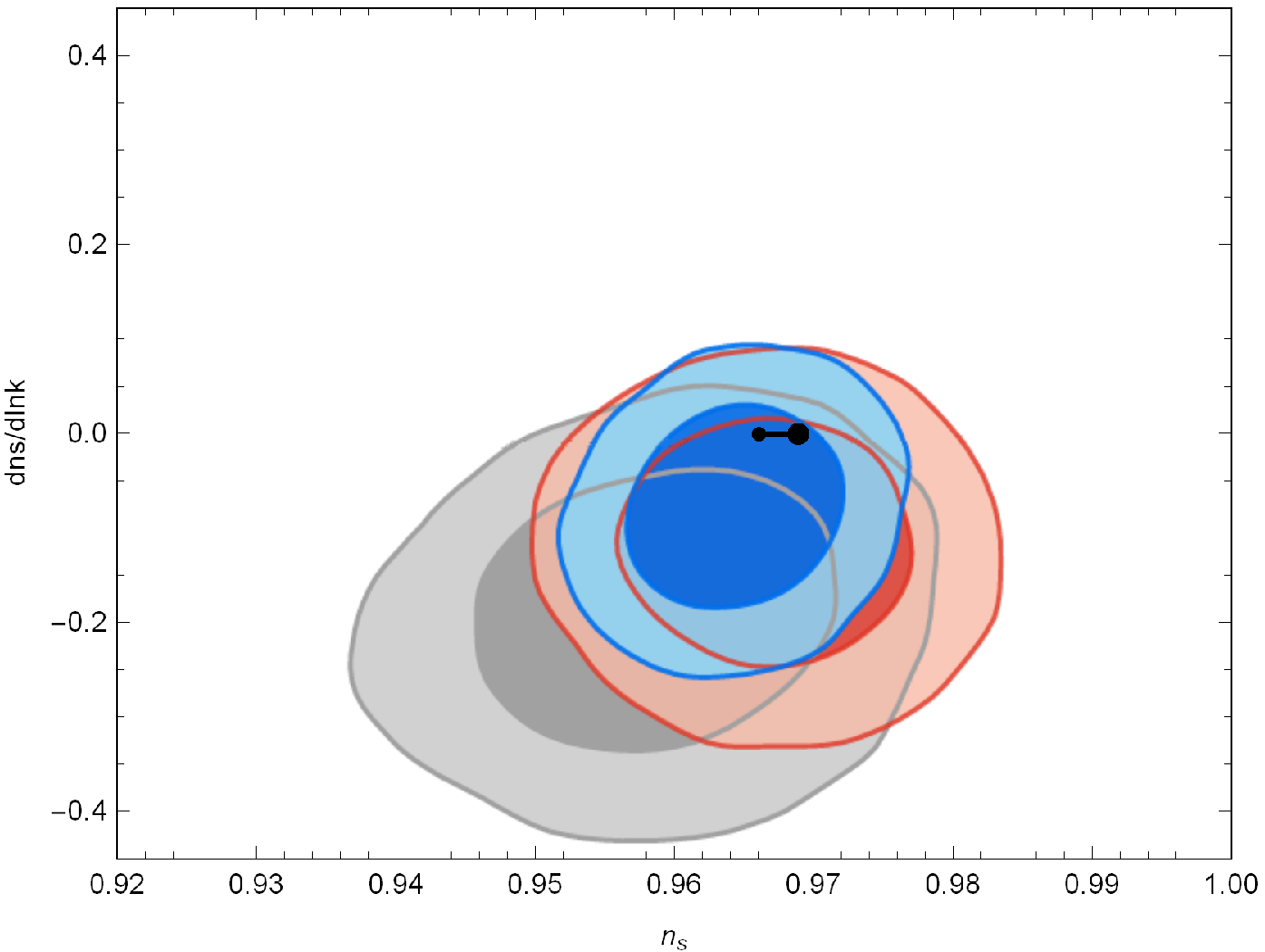}}
 \subfigure[$\alpha_s-n_s (n=2)$]{\includegraphics[width=4.3cm]{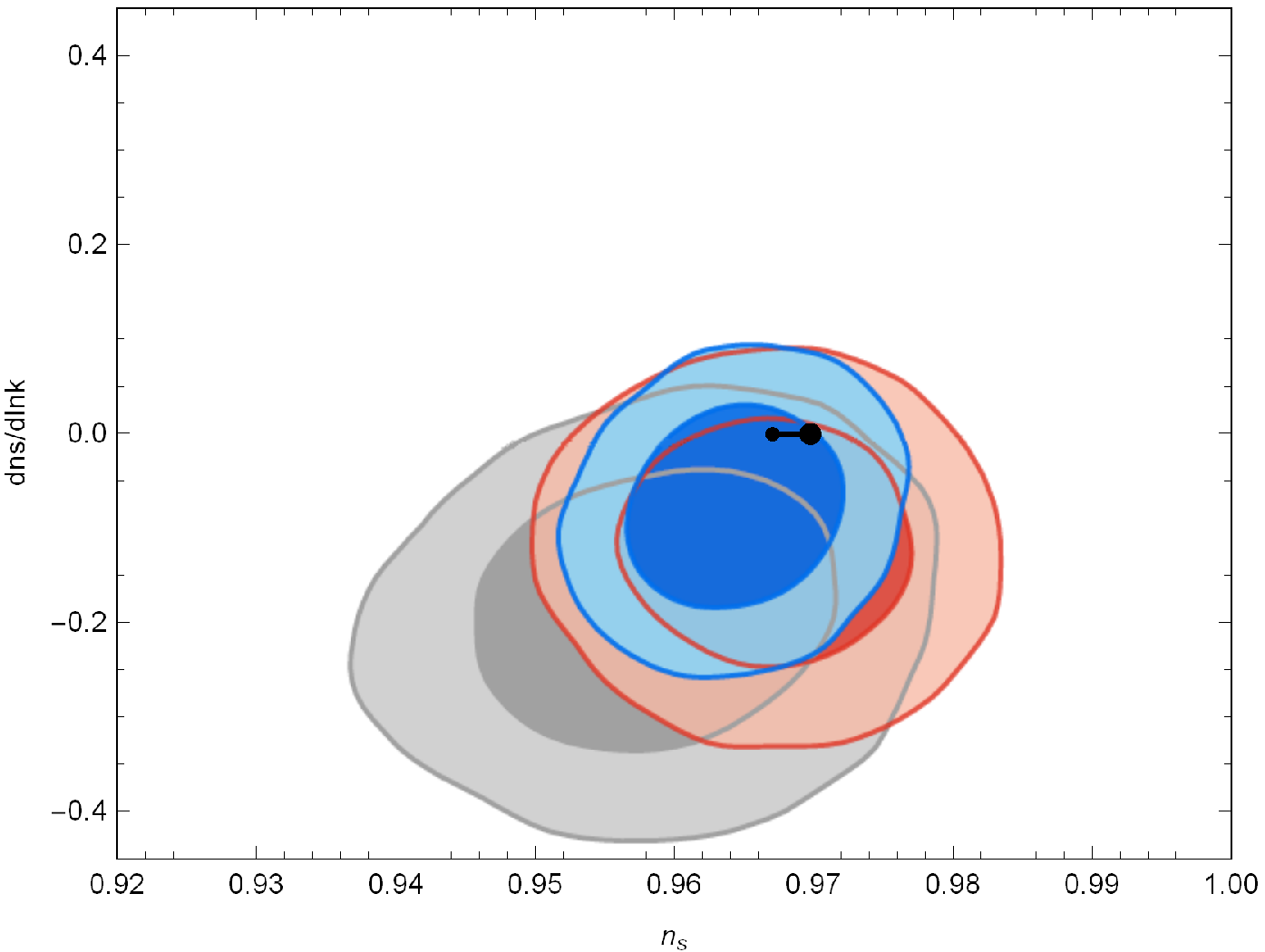}}
 \caption{$r-n_s$ for a) $n=0.5$, b) $n=1$, c) $n=2$; and $\alpha_s-n_s$ for d) $n=0.5$, e) $n=1$, f) $n=2$ diagrams of the model are compared with Planck data so that the small point belongs to $N=55$ and the large point corresponds to $N=60$. }\label{rns02}
\end{figure}

The dissipation coefficient is taken as a function of scalar field, $\Gamma = \Gamma_0 \phi^m$, and bulk viscous coefficient remains constant. Taking $\epsilon=1$ as a sign for end of inflation, and using Eq.(\ref{efold}), the scalar field at horizon crossing is given by
\begin{equation*}
\phi_\ast^{\nu} = \phi_e^{\nu} \left( 1+ {\nu \over n}\; N\right), \quad \phi_e^{\nu} = {6n^2 H_0 \over \Gamma_0}.
\end{equation*}
where $\nu = m-n+2$. Then, the slow-rolling parameters at this point come to
\begin{equation*}
\epsilon^\ast = {n \over n + (\nu)N}, \quad \eta^\ast = {2(n-1) \over n}\; \epsilon^\ast, \quad \sigma^\ast = {(n-2) \over 2n}\; \eta^\ast
\end{equation*}
\begin{equation*}
\beta^\ast = {m \over n}\; \epsilon^\ast, \quad \delta^\ast = {m(m-1) \over n^2}\; \epsilon^\ast {}^2.
\end{equation*}
The last term on the r.h.s of the scalar spectral index (\ref{strongns}) might be estimated by taking into account some good approximations. Since we are considering the slow-rolling inflation where the scalar field is the dominant component in our interest era, one could take $H^2 \gg \rho, \Pi$ (it is assumed that the energy density of fluid and its bulk viscosity are almost as same order; refer to \cite{cid}). Then, it could be extracted that
\begin{equation*}
{4 H' \over Q H} \; \bar{\chi} \simeq -2\beta^\ast - {3 \over 2} \epsilon^\ast = - {4m+3n \over 2n}\; \epsilon^\ast .
\end{equation*}
Substituting it in Eq.(\ref{strongns}), the scalar spectral index could be extracted in term of $n$ and $m$. Depicting the parameter, one could approximately arrives at the best values of $(n,m)$, see Fig.\ref{ns2}.\\
Staying in strong dissipative regime lead one to the following inequality
\begin{equation}\label{Q02}
Q={\Gamma \over 3H}>0 \quad \rightarrow \quad \Gamma_0 > 3H_0 \phi^{n-m},
\end{equation}
that should be satisfied always during the inflationary times. Therefore, $\Gamma_0$ should be bigger that the maximum of right hand side of above relation. On the other hand, from (\ref{phidot}), one could realize that for $n>0$, the scalar field has a decreasing behavior during the inflation. Now, if $n-m>0$ the maximum of r.h.s is obtained as $3H_0 \phi_\ast^{n-m}$, and if $n-m<0$ the maximum is obtained as $3H_0 \phi_e^{n-m}$. Then, one can write Eq.(\ref{Q02}) as follows
\begin{equation}\label{QH02}
\Gamma_0 = \Theta \; \Sigma \; H_0,
\end{equation}
where
\begin{eqnarray*}
\Sigma & = &  3^{\nu/2} \Big( 6n^2 \bar{N} \Big)^{n-m \over 2},  \\
\Theta & = & \theta^{\nu \over 2}; \qquad \bar{N} \equiv \left\{  \begin{array}{ll}
                                                            1 + {\nu \over n}\; N & \text{if} \quad n-m>0, \\
                                                            1                     & \text{if} \quad n-m<0,
                                                          \end{array} \right.
\end{eqnarray*}
and the parameter $\theta$, which is the constant of proportionality in Eq.(\ref{Q02}), is much bigger than one to satisfy the condition $Q \gg 1$. The other important condition is $T_r>H$ which is an essential feature of warm inflation. if the fluid energy density is taken as $\rho=C_\gamma T_r^4$ \cite{cid}, from Eqs.(\ref{radiation}) and (\ref{phidot}) there is
\begin{equation}
T_r = \left( {1 \over \gamma C_\gamma} \left[ {12 n^2 H_0^3 \over \Gamma_0}\; \phi^{3n-m-2} + 3\xi_0 H_0 \phi^n \right] \right)^{1 \over 4}.
\end{equation}
Then, the mentioned condition comes to the following expression which hold during inflation as
\begin{equation}\label{TH02}
(C_\gamma \gamma \phi^{3n}) H_0 ^3 - \left( {12n^2 \over \Theta \Sigma} \; \phi^{2n-m-2} \right) H_0 - 3\xi_0 <0.
\end{equation}
On the other hand, to meet the latest observational data, the free parameters of the model are picked out in order to put the model predictions about the perturbation parameters in acceptable range given by Planck data. In this regards, we are to consider two important perturbation relations that could help to constrain the free parameters of the model, namely the amplitude of scalar perturbation and the tensor-to-scalar ration. According to \cite{taylor}, it was mentioned that the tensor perturbation do not coupled strongly to the thermal background and so the gravitational waves are only produced by quantum fluctuations. On the other side, following \cite{Bhattacharya} one realizes that there could be an extra temperature dependence term in the amplitude of tensor perturbation. Considering both results leads one to the following upper bound for the model parameter $H_0$ as
\begin{eqnarray}\label{Hr02}
H_0 & < & \left( {\Theta \; \Sigma \over 6n^2 \bar{N}} \right)^{n \over \nu} \sqrt{ 2\pi^2 r^\ast \mathcal{P}_s^\ast }, \nonumber\\
H_0 & < & \left( {\Theta \; \Sigma \over 6n^2 \bar{N}} \right)^{n \over \nu} \sqrt{ 2\pi^2 r^\ast \mathcal{P}_s^\ast \over \coth\big(k/2T \big) },
\end{eqnarray}
where $r^\ast$ and $\mathcal{P}_s^\ast$ are respectively the tensor-to-scalar ratio and the amplitude of scalar perturbation at horizon crossing and are given by $r^\ast < 0.11$ and $\ln\Big( 10^{10} \mathcal{P}_s \Big) = 3.062$, according to Planck data. Expressing the amplitude of scalar perturbation in term of $H_0$, one can use Eq.(\ref{QH02}) to get a proper value for $H_0$. The amplitude of scalar perturbation takes a complicated form, and getting analytical solution is not easy. Therefore, we plot $\mathcal{P}_s$ in term of $H_0$ in Fig.\ref{psG}, and it could be utilized for obtaining a proper value for $H_0$.
To reach the predicted amplitude of scalar perturbation by Planck, the parameter $H_0$ should be about $H_0 = 5.6 \times 10^{-7}$, $6.4 \times 10^{-6}$, and $4.6 \times 10^{-4}$ respectively for $n=0.5$, $1$ and $2$, where the other constants are taken as:$\gamma=1.5$, $m=3$, $N=55$, $\theta=2\time 10^{10}$ $\xi_0=7 \times 10^{-14}$ ($M_p^3$), $k=0.002$ ($\rm{Mpc}^{-1}$) and $T_r=T=2 \times 10^{-5}$ ($M_p$). Plugging this value into above inequalities Eqs.(\ref{TH02}) and (\ref{Hr02}), it is easily verified that all the conditions are satisfied for the obtained value of $H_0$. \\
Considering the potential and also ratio of the energy densities might be an examination whether the parameters are selected carefully. Fig.\ref{pot2} displays the potential behavior versus scalar field in inflationary times for different $n$ as $n=0.5, 1, 2$. It is clear that for all cases, the potential stands below the Planck energy scale. In addition, the scalar field rolls down slowly from the top to the minimum of its potential and the inflation could last enough.\\
The other test is ratio of energy densities that Fig.\ref{energya} makes it easy for us. In contrast to the previous case, the fluid energy density is quite below the scalar field energy density which in turn states that the free parameters have been picked out properly. \\
To complete the work in this section, one could plot $r-n_s$ and $\alpha_s-n_s$ diagrams to check out, how the model prediction could come close to $68\%$ CL area of Planck data. Fig.\ref{rns02} illustrates both diagrams, where it could be found out that the result could stay in $68\%$ CL area implying another argument for capability of the model.

\subsection{Typical Example 3: $\Gamma := \Gamma_0 H^2$ and $\xi=\xi_0 \rho$}
\begin{figure}
 \centering
 \includegraphics[width=6cm]{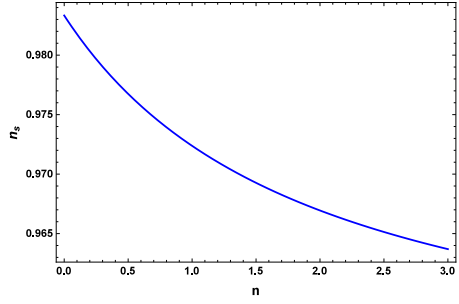}
 \caption{$n_s$ versus $n$.}\label{ns3}
\end{figure}

\begin{figure}[h]
 \centering
 \subfigure[$n=1$]{\includegraphics[width=4.3cm]{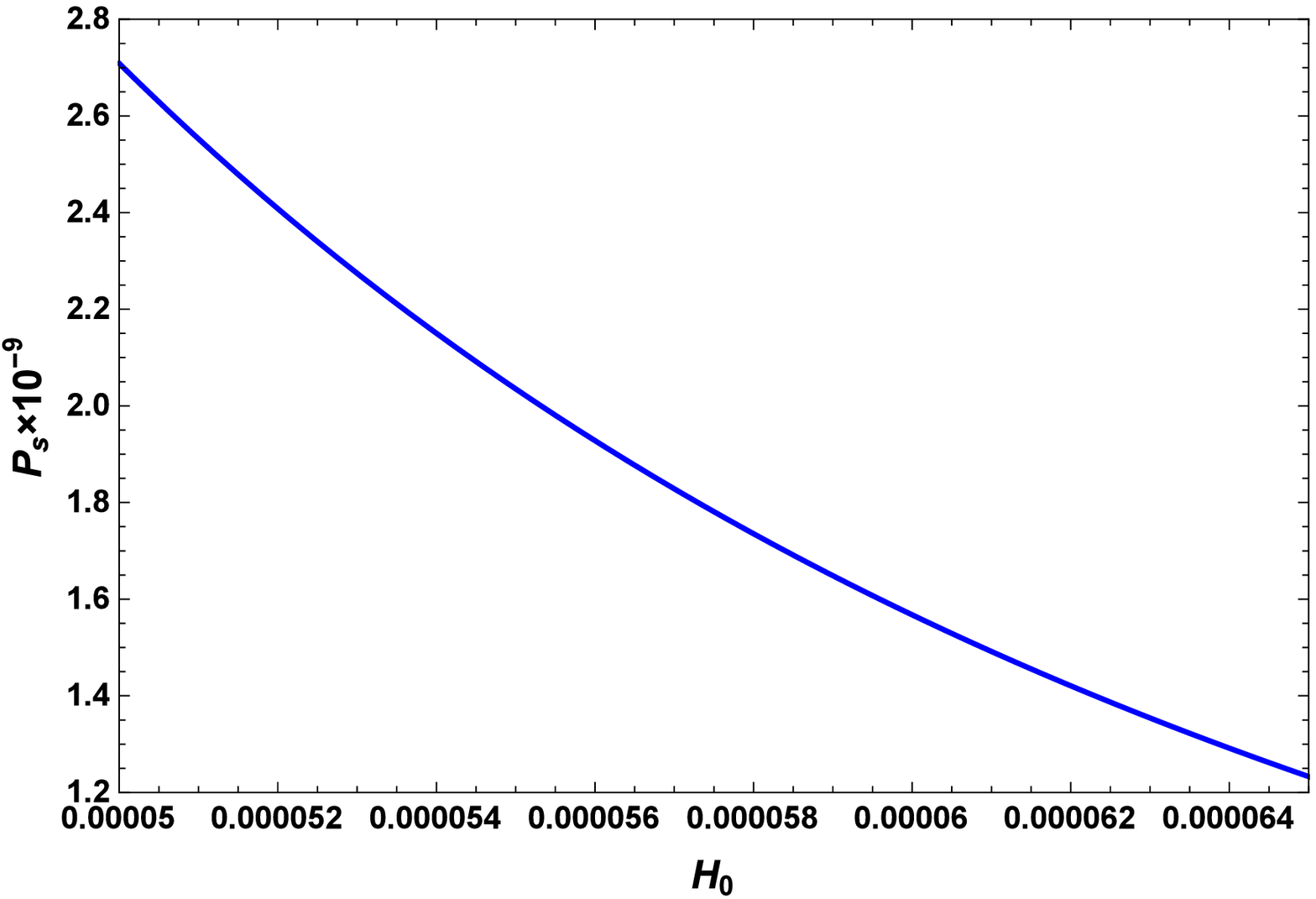}}
 \subfigure[$n=2$]{\includegraphics[width=4.3cm]{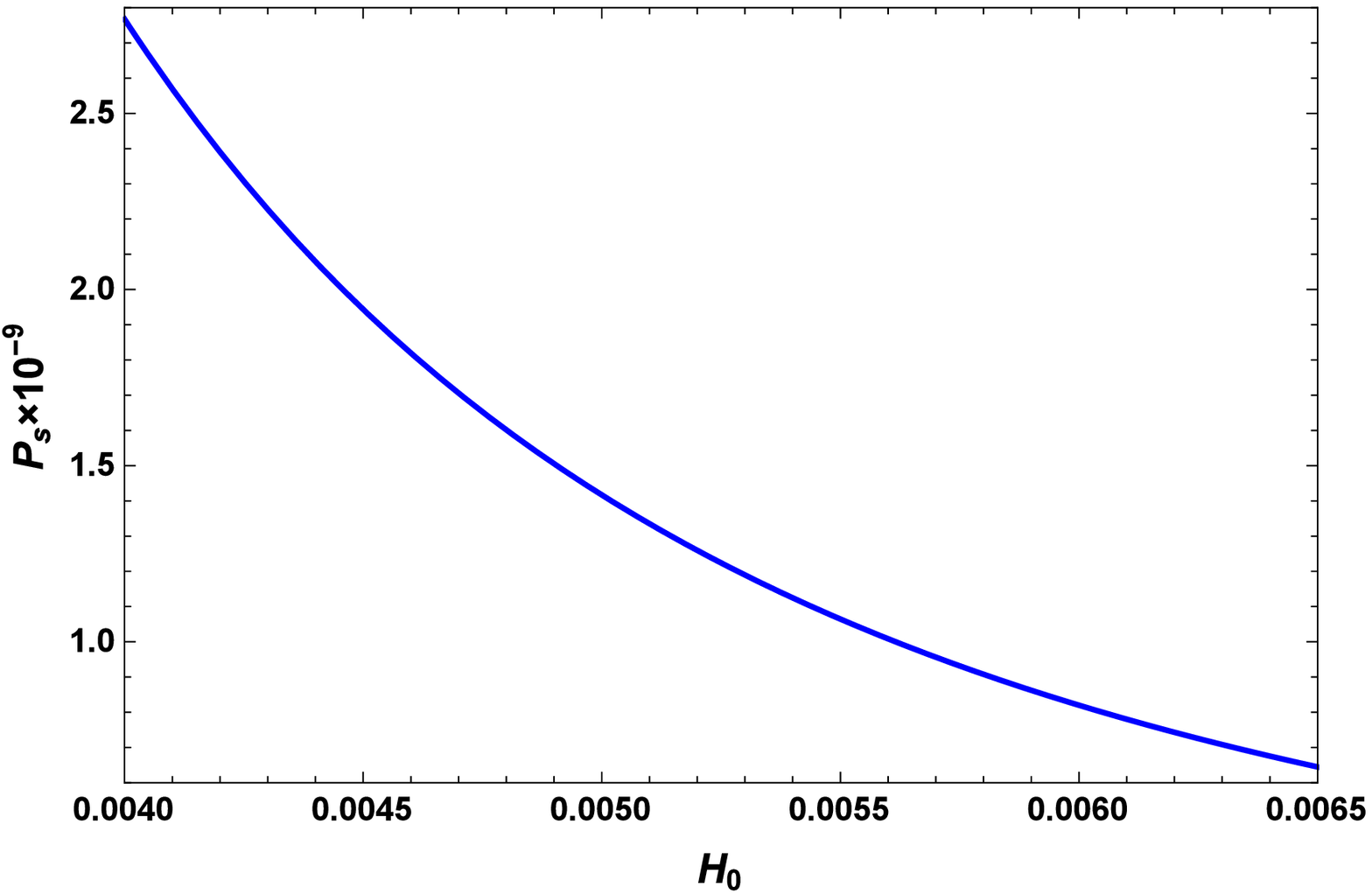}}
 \caption{Amplitude of scalar perturbation in term of the constant parameter $H_0$ is plotted where the constant parameters are: $\gamma=1.5$, $N=60$, $\xi_0=3 \times 10^{-14}$,$k=0.002$ and $T_r=T=2 \times 10^{-5}$ and a)$\theta=1\times 10^{20}$ b)$\theta=1\times 10^{10}$. }\label{psgamma}
\end{figure}

\begin{figure}[h]
 \centering
 \subfigure[$n=1$]{\includegraphics[width=4.3cm]{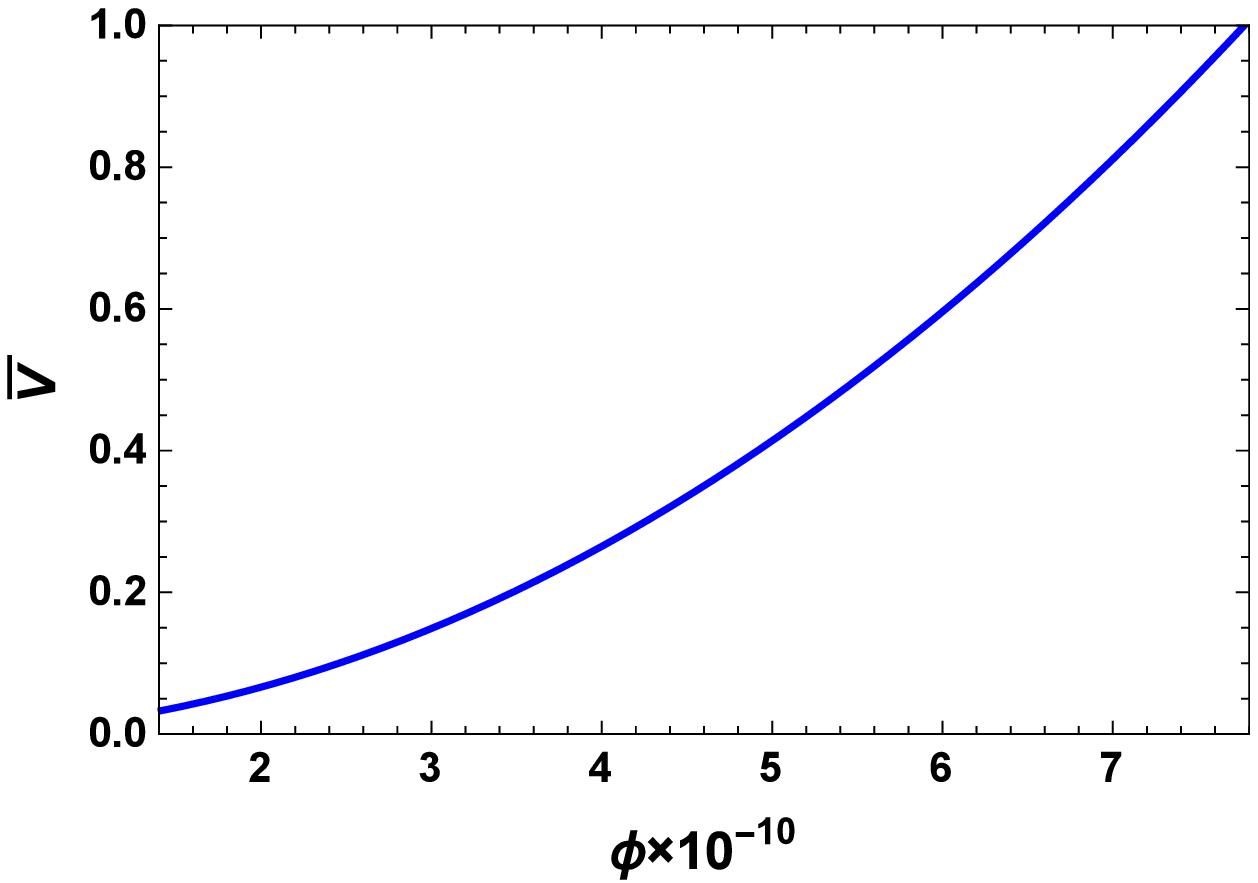}}
 \subfigure[$n=2$]{\includegraphics[width=4.3cm]{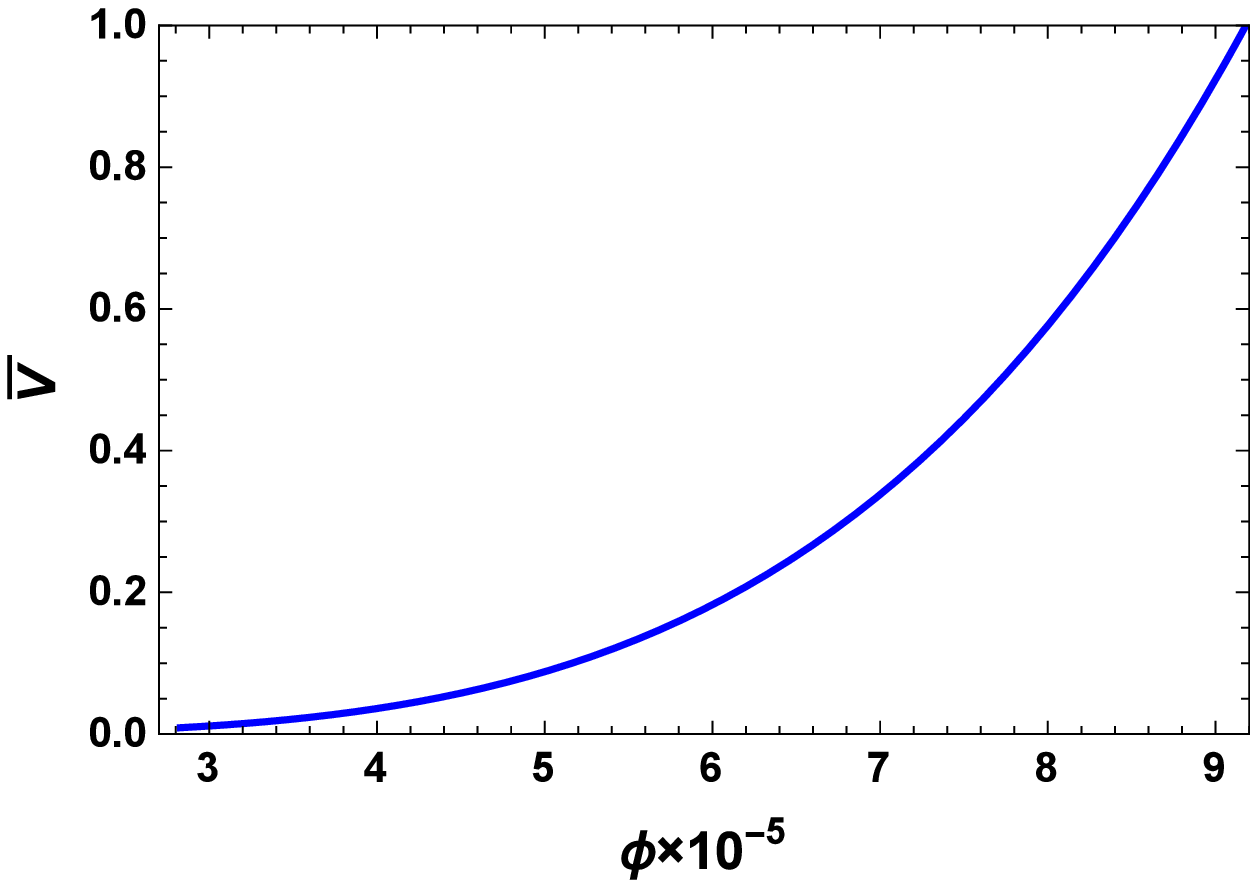}}
 \caption{The potential versus scalar field during the inflationary times for the constant parameters: $\gamma=1.5$, $N=60$, $\xi_0=3 \times 10^{-14}$,$k=0.002$ and $T_r=T=2 \times 10^{-5}$ and a)$\theta=1\times 10^{20}$ b)$\theta=1\times 10^{10}$. The parameter $\bar{V}$ is defined by $\bar{V}=V/V_i$ where $V_i$ is the initial value of the potential.}\label{pot02}
\end{figure}

\begin{figure}[h]
 \centering
 \subfigure[$n=1$]{\includegraphics[width=4.3cm]{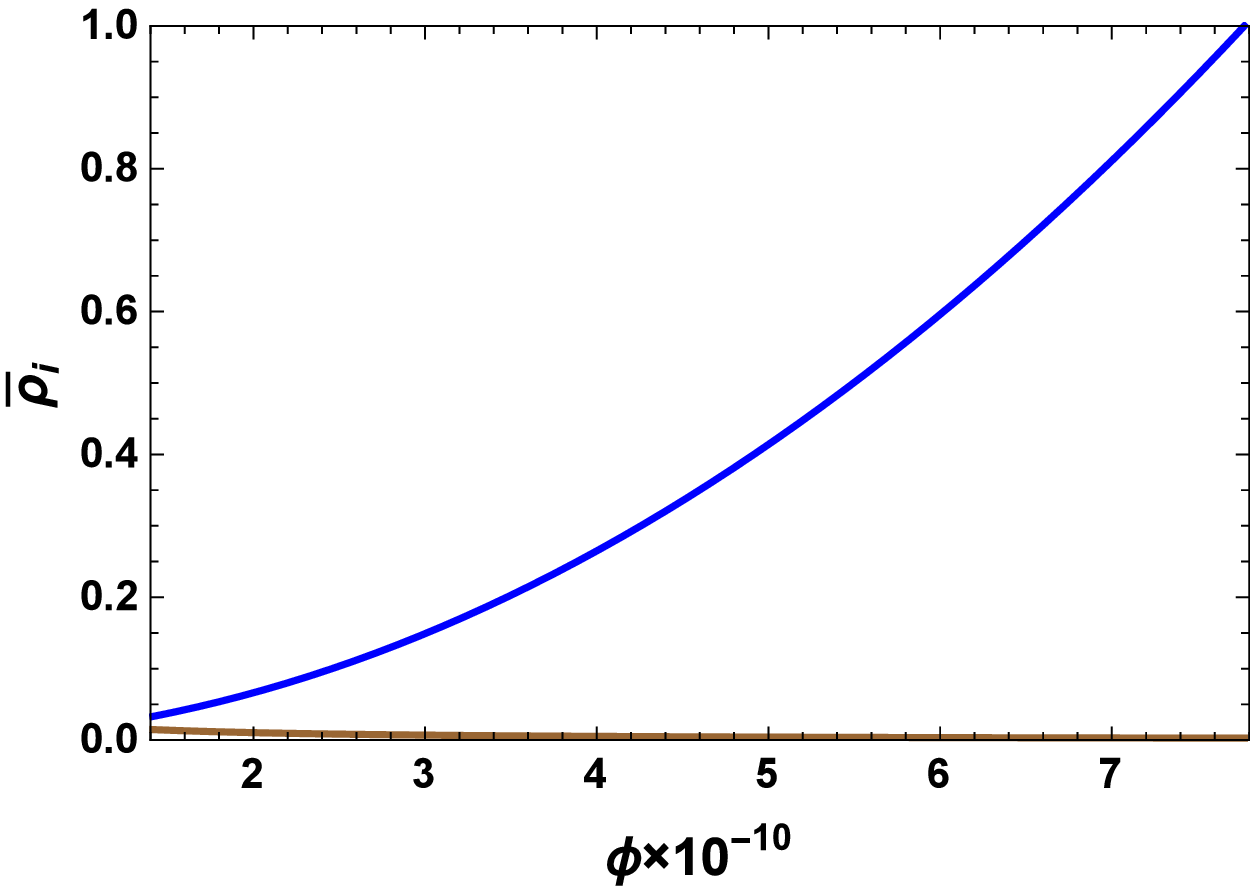}}
 \subfigure[$n=2$]{\includegraphics[width=4.3cm]{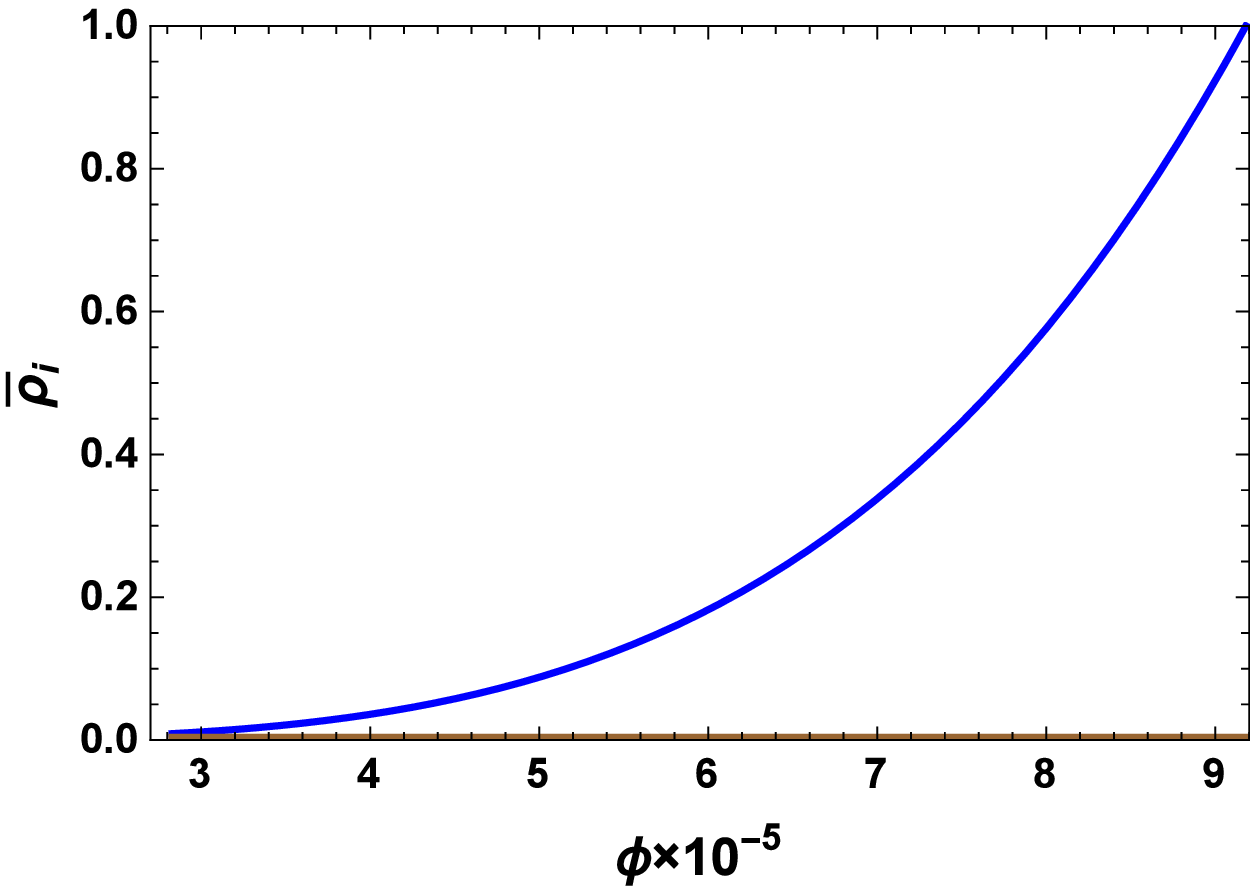}}
 \caption{Energy density of fluids versus scalar field during the inflationary times for the constant parameters: $\gamma=1.5$, $N=60$, $\xi_0=3 \times 10^{-14}$,$k=0.002$ and $T_r=T=2 \times 10^{-5}$ and a)$\theta=1\times 10^{20}$ b)$\theta=1\times 10^{10}$. The parameter $\bar{\rho}_i$ (the subscript "i" stands for scalar field and matter) is defined by $\bar{\rho}_i=\rho_i/\rho_0$ where $\rho_0$ is the initial value of the scalar field energy density.}\label{energy02}
\end{figure}

\begin{figure}[h]
 \centering
 \subfigure[$r-n_s (n=1)$]{\includegraphics[width=4.3cm]{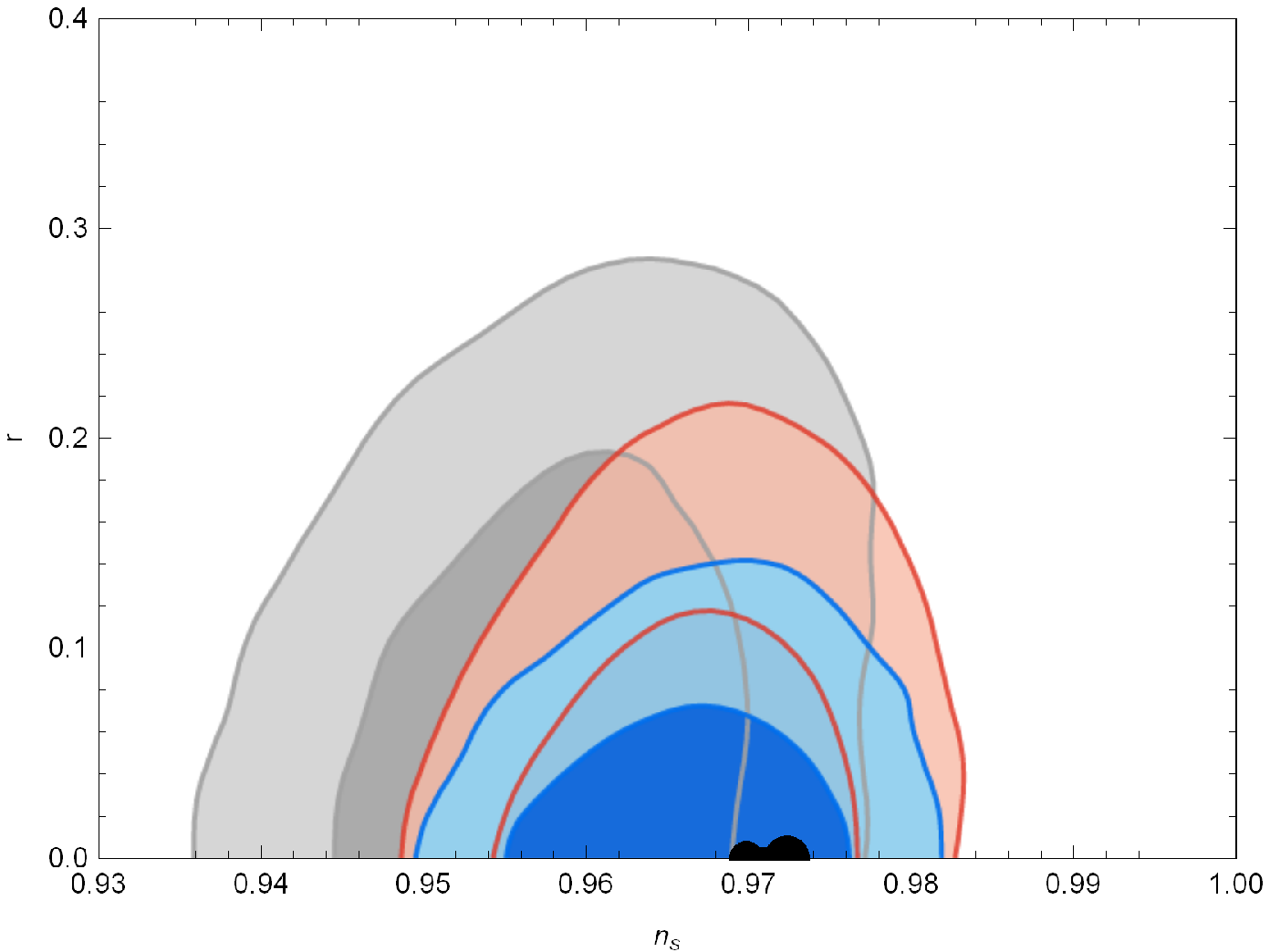}} \hspace{1cm}
 \subfigure[$r-n_s (n=2)$]{\includegraphics[width=4.3cm]{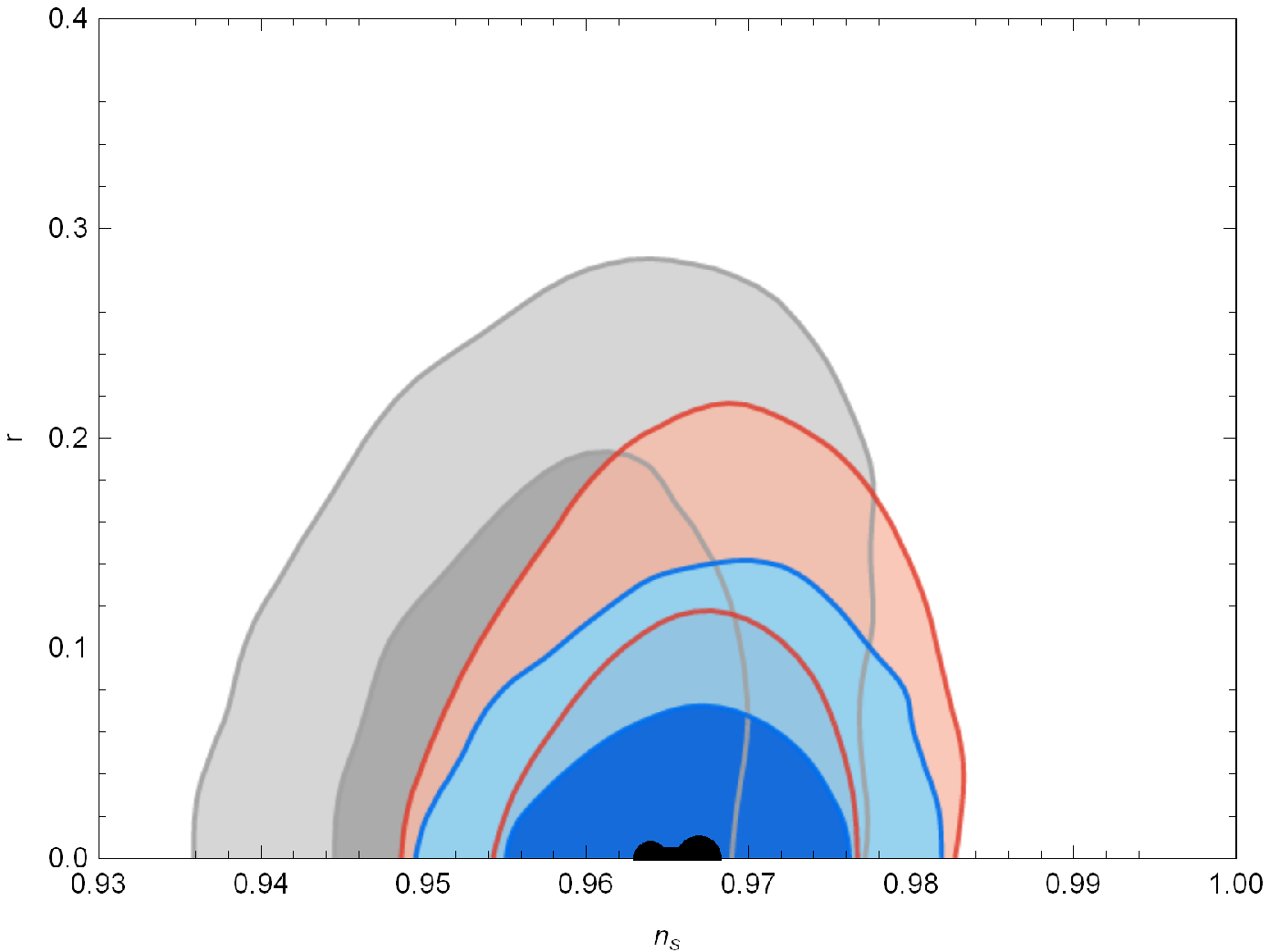}}
 \subfigure[$\alpha_s-n_s (n=1)$]{\includegraphics[width=4.3cm]{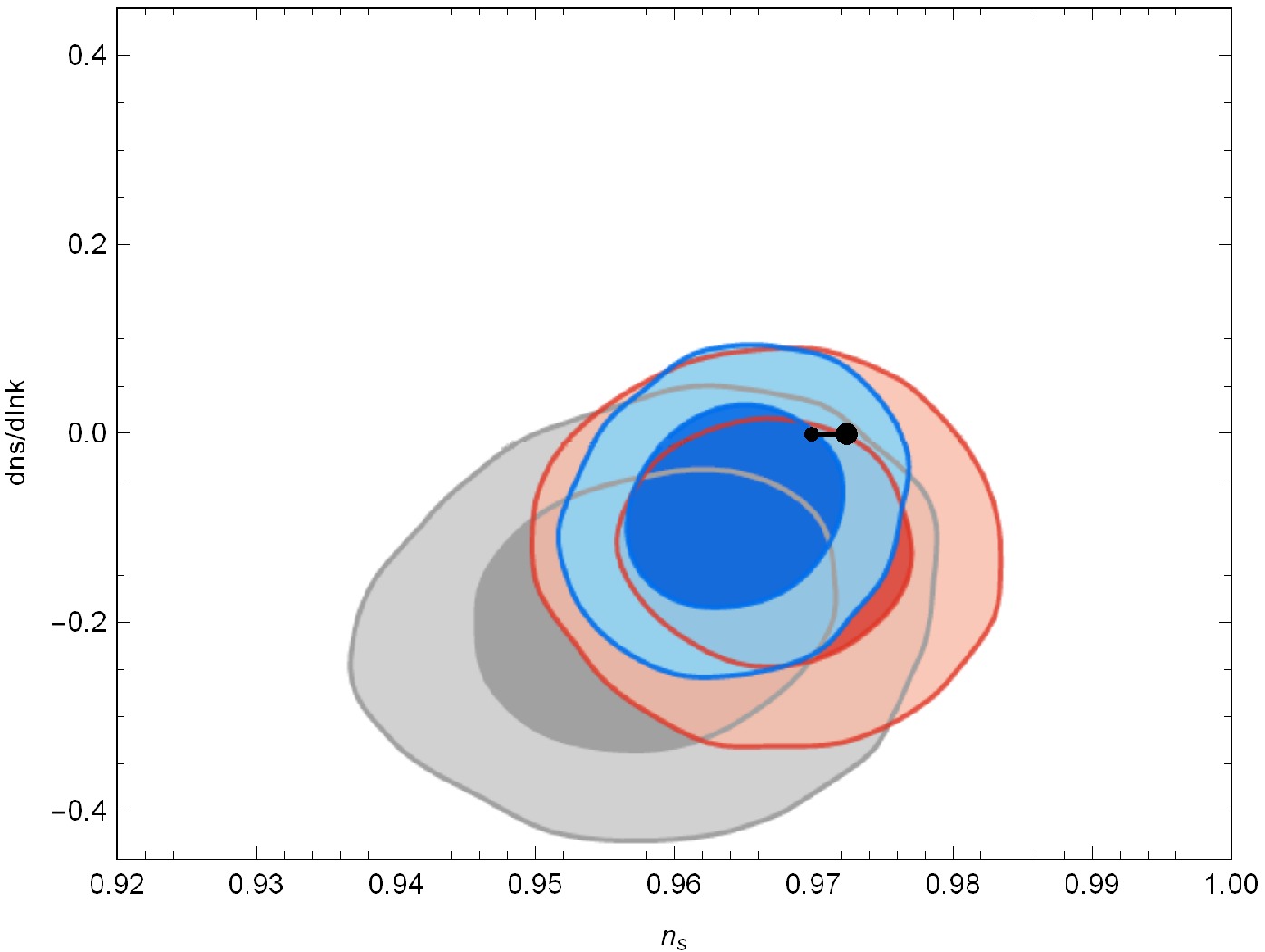}} \hspace{1cm}
 \subfigure[$\alpha_s-n_s (n=2)$]{\includegraphics[width=4.3cm]{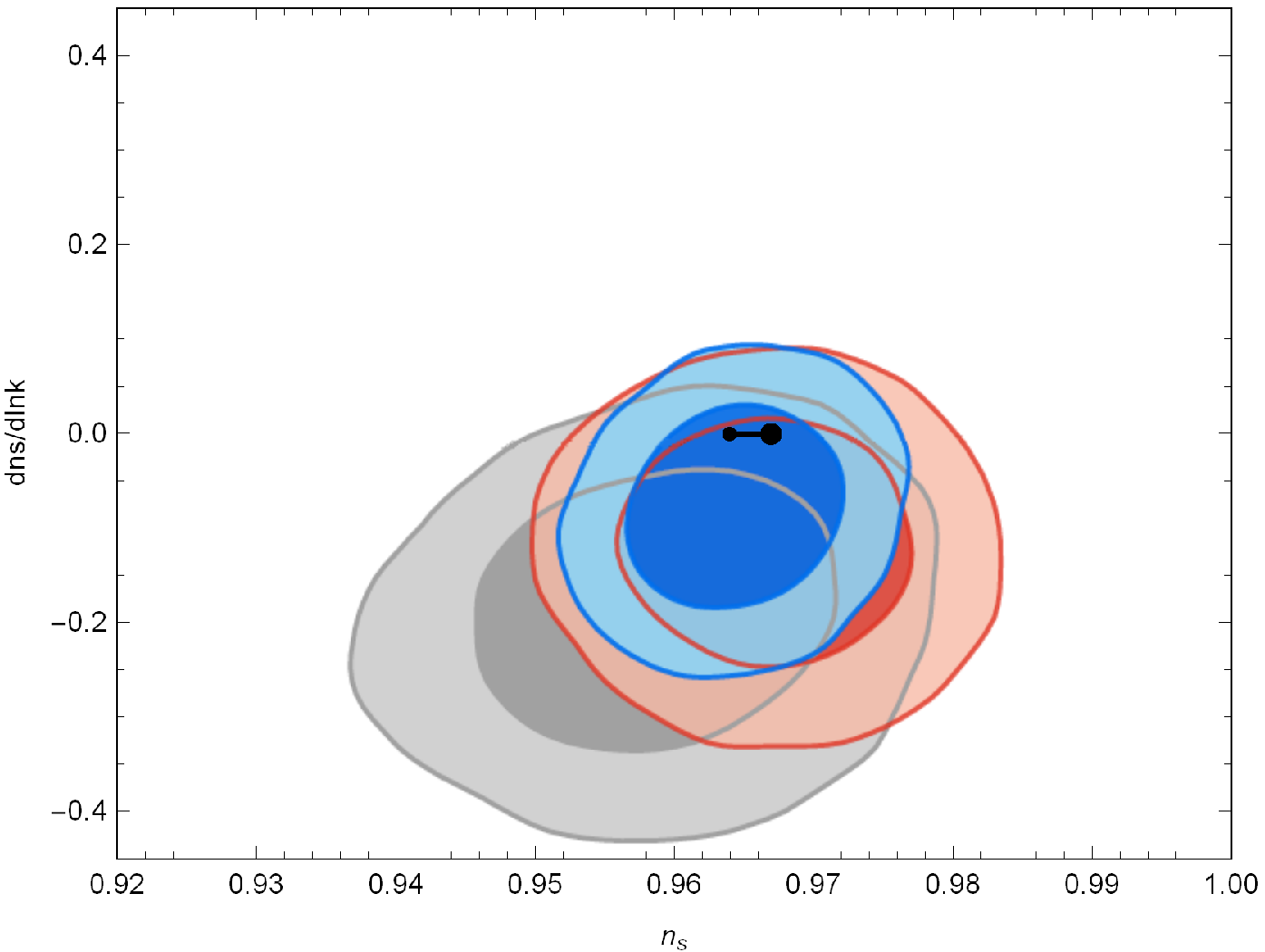}}
 \caption{$r-n_s$ for a) $n=1$, b) $n=2$; and $\alpha_s-n_s$ for c) $n=1$, d) $n=2$ diagrams of the model are compared with Planck data so that the small point belongs to $N=55$ and the large point corresponds to $N=60$. }\label{rns03}
\end{figure}

Here we take a more general view on the case, and assume the coefficients as variable in which the dissipation coefficient is taken proportional to the Hubble parameter $\Gamma = \Gamma_0 H^2$, and the bulk viscous coefficient is taken as $\xi = \xi_0 \rho$, where $\Gamma_0$ and $\xi_0$ are constant. \\
At the end of inflation, the slow-rolling parameter $\epsilon$ reaches one, and using the number of e-fold relation (\ref{efold}), the perturbations exit horizon as the scalar field arrives at
\begin{equation}\label{phii2}
  \phi_\ast = \phi_e \Big[ {N(2+n) \over n} + 1 \Big]^{1 \over n+2}, \qquad \phi_e =\left( {\frac{6n^2}{\Gamma_{0} H_{0}}} \right)^{1 \over n+2}.
\end{equation}
Therefore, the slow-rolling parameters are estimated at horizon crossing as
\begin{equation*}
\epsilon^\ast = {n \over n + (2+n)N}, \quad \eta^\ast = {2(n-1) \over n}\; \epsilon^\ast, \quad \sigma^\ast = {(n-2) \over 2n}\; \eta^\ast,
\end{equation*}
\begin{equation*}
\beta^\ast = 2 \epsilon^\ast, \quad \delta^\ast = {2n+2(n-1) \over n} \epsilon^\ast {}^2.
\end{equation*}
Following the same process and using the same reasons as the previous cases, the last term of Eq.(\ref{ns}) could be estimated as
\begin{equation*}
{4 H' \over Q H} \; \bar{\chi} \simeq -2\beta^\ast - {3 \over 2} \epsilon^\ast = - {11 \over 2}\; \epsilon^\ast.
\end{equation*}
Then, the scalar spectral index is extracted in terms of number of e-fold and $n$, where Fig.\ref{ns3} determines its behavior of $n_s$ versus $n$ for a given number of e-folds, $N=60$. Checking the plot, it is seen that for $n=0.5$ the scalar spectral index is out of observational range, then we ignore this case for the rest of this subsection. \\
Strong dissipation regime leads to the following inequality between the free parameters $H_0$ and $\Gamma_0$
\begin{equation}\label{Q03}
\Gamma_0 > {3 \over H_0 \phi^n}.
\end{equation}
During the inflation, $\Gamma_0$ should be bigger than the maximum of the r.h.s of relation to stay in strong dissipative regime. Since for $n>0$ the scalar field has a decreasing behavior, according to Eq.(\ref{phidot}), the maximum is obtained for $\phi_e$, and by introducing a constant parameter $\theta (\gg 1)$, one arrives at
\begin{equation}\label{QH03}
\Gamma_0 = \Theta \; \Sigma \; H_0^{-1}
\end{equation}
where
\begin{eqnarray*}
\Sigma & = &  {3^{n+2/2} \over \Big( 6n^2 \bar{N} \Big)^{n/2} },   \\
\Theta & = & \theta^{n+2 \over 2}, \qquad \bar{N} \equiv \left\{  \begin{array}{lll}
                                                            1                      & \text{if} &  n>0, \\
                                                            1 + {n+2 \over n}\; N  & \text{if} &  n<0,
                                                          \end{array} \right.
\end{eqnarray*}
The temperature of thermal bath is stated as $\rho = C_\gamma T^4$ \cite{cid}. From Eqs.(\ref{radiation}) and (\ref{phidot}), we have
\begin{equation}
T_r = \left( {12 n^2 H_0 \over C_\gamma \Gamma_0 \big( \gamma - 3\xi_0 H_0 \phi^n \big)} \; \phi^{n-2} \right)^{1 \over 4}.
\end{equation}\label{TH03}
Then the condition for dominating thermal fluctuation over the quantum fluctuation, namely $T_r>H$, is given by
\begin{equation}
\Big( 3\xi_0 \phi^{4n+2} \Big) H_0^3 - \Big( \gamma \phi^{3n+2} \Big) H_0^2 + {12n^2 \over \gamma C_\gamma \Theta \Sigma} >0.
\end{equation}
Other condition for the parameter $H_0$ is obtained by considering the tensor-to-scalar ratio which is bounded as $r^\ast < 0.11$ \cite{planck}. Following \cite{taylor} and \cite{Bhattacharya} two different expressions for the amplitude of tensor perturbation are found where for each case $H_0$ satisfies the following constraint, respectively
\begin{eqnarray}\label{Hr03}
H_0 & < & \left( {\Theta \Sigma \over 6n^2 \bar{N}} \right)^{n \over n+2} \sqrt{ 2\pi^2 r^\ast \mathcal{P}_s^\ast }, \\
H_0 & < & \left( {\Theta \Sigma \over 6n^2 \bar{N}} \right)^{n \over n+2} \sqrt{ 2\pi^2 r^\ast \mathcal{P}_s^\ast \over \coth\big(k/2T \big) }.
\end{eqnarray}
However, by utilizing the amplitude of scalar perturbation one could restrict $H_0$ properly since there is an exact value for $\mathcal{P}_s$ \cite{planck}. Using (\ref{QH03}), the amplitude of scalar perturbation could be derived in terms of $H_0$. Since getting an analytical solution is complicated, one could depict $\mathcal{P}_s$ in term of $H_0$ as in Fig.\ref{psgamma}, and try to read the proper value for $H_0$, so that for $n=1$ and $n=2$ the parameter is estimated respectively as $H_0=9.80 \times 10^{-5}$ and $H_0=1.47 \times 10^{-2}$ by taking $\gamma=1.5$, $N=60$, $\xi_0=3 \times 10^{-14}$ ($M_p^{-1}$),$k=0.002$ ($\rm{Mpc}^{-1}$) and $T_r=T=2 \times 10^{-5}$ ($M_p$) and a)$\theta=1\times 10^{20}$ b)$\theta=1\times 10^{10}$. For the mentioned values of the parameters, it could be checked that the conditions (\ref{TH03}) and (\ref{Hr03}) could be satisfied for the acquired value of $H_0$, that indicates that the model could properly describe warm inflationary scenario. \\
The potential is illustrated versus scalar field during the inflation as in Fig.\ref{pot02} so that it has the same behavior as the previous case where the scalar field rolls down slowly toward the minimum of potential and let inflation last enough. \\
Behavior of the energy density of scalar field and fluid could be compared based on Fig.\ref{energy02}, where at the initial of inflation the scalar field is the dominant component. By passing time, it deceases and both come close to each other at the end of inflation.\\
Capability of the model could be examined by exhibiting $r-n_s$ and $\alpha_s-n_s$ diagrams as in Fig.\ref{rns03}. The diagrams show that for both chosen values of $n$, the result stands in $68\%$ CL, pointing out that the case could be an acceptable case.  \\

\section{Conclusions}\label{conclusion}
Using the Hamilton-Jacobi formalism, we considered viscous warm inflation. After deriving the general dynamical equations and the necessary perturbation parameters, we investigated the model for three different cases in the strong dissipative regime.\\
In the first typical example, which is a simple case, the dissipation and bulk viscous coefficients are both taken as constants. It is explained that this case come to undesirable result and could not describe the scenario of warm inflation precisely.   \\
Generalizing the first case, the dissipation coefficient is taken as a function of scalar field the second typical example. Then warm inflation conditions are studied that lead to some bounds for the model parameters. On the other hand, from the observational data there is an exact value for the amplitude of scalar perturbation that is used to restrict the parameter $H_0$ exactly. Then, the amplitude of scalar perturbation is depicted versus $H_0$, where we found out best value for $H_0$ that properly satisfies all other conditions of warm inflationary scenario. In order to examine our result, we plotted the potential and energy densities which show that they all stayed below the Planck energy density scale. Also the smallness of fluid energy density against scalar field clearly is displayed. Further examinations are performed by analyzing $r-n_s$ and $\alpha_s-n_s$ diagrams, where both determine that the model predictions place in $68\%$ CL area.  \\
A more general situation was picked out as the last typical example where dissipation coefficient depends on scalar field and the bulk viscous coefficient is a function of fluid energy density. The same process as the previous case is followed where the amplitude of scalar perturbation is depicted in terms of $H_0$ to extract a proper value for it. For the chosen parameters, the potential and the energy densities are illustrated and the result of it show that the scalar field energy density and the fluid energy density have compatible behavior with our assumption. More examinations for the model are performed by $r-n_s$ and $\alpha_s-n_s$ diagrams, when we realized that it stands in $68\%$ CL area indicating the consistency between model and Planck data. \\
Noticing the above conclusion, comparison with presented literature in the introduction part is stated briefly. For the presented work, the corresponding potential for the given Hubble parameter is expressed in Eq.(\ref{pot}), where for $n=0.5, 1, 2$ there is approximately linear, quadratic, and quartic potential. Although the first case could not stand as a candidate for the warm inflation, the second case, where dissipation coefficient is taken as a function of scalar field and coefficient of bulk viscosity is constant, establishes a good candidate for warm inflation for three type of the potential. On the other hand the linear potential for the third example could not lead to good result. However, the quadratic and quartic potentials are more desirable, and give our intended outcome. Then, in comparison to \cite{delCampo}, the running of scalar spectral index is obtained in observational range. Also, compared to \cite{sharif}, all the result are acquired for enough number of e-fold.










\end{document}